# A study on the leverage effect on financial series using a TAR model: A Bayesian approach

**Oscar Espinosa M.Sc.**

Director of the Researching Group of Economic Models and Quantitative Methods (IMEMC),

Universidad Nacional de Colombia. ORCID: 0000-0003-4893-0509

**Fabio Nieto Ph.D.**

Tenured Professor, Universidad Nacional de Colombia.

E-mail: fhnietos@unal.edu.co

**Abstract.** This research shows that under certain mathematical conditions, a threshold autoregressive model (TAR) can represent the leverage effect based on its conditional variance function. Furthermore, the analytical expressions for the third and fourth moment of the TAR model are obtained when it is weakly stationary. This research makes an empirical application, where TAR model is fitted using Nieto's (2005) methodology and VAR-GARCH multivariate model is estimated through A-BEKK approach to the BOVESPA stock index. Finally, both statistical models are compared, via conditional and unconditional moments, and the representation of the leverage effect.

**Financial Support**

This research was financed by the Science, Technology and Innovation Administrative Department of Colombia (Colciencias) and the National University of Colombia (Bogota), by means of the Resolution 1265 of 2017, within the *Young Researchers Program 2016* (call 761).

**Acknowledgments**

The authors expresses his gratitude to Michael McAleer Ph.D., Luc Bauwens Ph.D., Manabu Asai Ph.D., Massimiliano Caporin Ph.D., Bruce Hansen Ph.D., Henry Olan Ph.D., Audrone Virbickaite Ph.D., Concepción Ausín Ph.D. Additionally, the authors expresses thanks to the attendees of the *XXVIII Statistics International Symposium* and of the *XII International Symposium Of Finance Experts*, for their suggestions and comments regarding various sections of this paper.

## 1. Introduction

In the stock market field, constructing models is vital in order to explain the evolution and dynamics of the financial time series, due to their specific characteristics, which differ from other variables of economic analysis. Commonly, these series present excess of kurtosis, slow decay of autocorrelation in square returns and absolute returns, high volatility periods frequently preceded by big negative returns and asymmetric behaviors (Granger y Ding, 1995; Cont, 2001; Sewell, 2011).

In the specialized literature the most frequent models used for financial series have been those derived from the ARCH (*Autoregressive Conditional Heteroscedasticity*) family, initiated by Engle (1982). These models posses the feature of modeling the volatility of financial returns through the conditional variance, depending on the past of its innovations (also called news). Since then, this family of models has been extended and sophisticated in order to obtain better explanations of the financial series.

However, other non-linear models have been created for these studies. Stochastic volatility models promulgated by Taylor (1986), the STAR models (*Smooth Threshold Autoregressive*) proposed by Teräsvirta and Anderson (1992), the SETAR models (*Self-Exciting Threshold Autoregressive*) studied by Tong (1990), the Hamilton's Regime-Switching model (1994) and the TAR (*Threshold Autoregressive*) analyzed by Tong (1978), Tong and Lim (1980), Tsay (1989) and Nieto (2005).

The latter regime shift models, although not designed for financial series analysis, as it is the case of the ARCH family; have permitted, in some cases, the capture of different patterns implicit in the historic behavior of time series.

Therefore, it is the aim of this paper to study analytically one of the features mentioned at the beginning of this section, commonly known as leverage effect. This feature enables the observation of an asymmetric response in the conditional variance of financial series facing negative profitability, different from the response given with positive profitability. This is one of the main reasons for the general increase of uncertainty towards stockbrokers when precipitous drops occur in the stock market (Black, 1976; Christie, 1982).

Moreno (2010), and Nieto and Moreno (2014) were the pioneers of TAR modeling for financial series (via conditional distribution), without studying specifically the leverage effect representation. An important characteristic of this is the fact that the conditional variance of past data explains certain heteroscedasticity in the return series, which shows that a typical trajectory of the threshold stochastic process can represent conglomerates of external values (Nieto and Moreno, 2014; 2016).

The document is divided in five sections. The first part is the introduction. The second section formally describes the TAR model and its conditional and non-conditional moments. In the following section new mathematic results for the TAR model are exposed, related with the analytic calculus of the third and fourth standardized moments, and deductions based on the conditional variance of the data, which explains a new type of leverage effect from a threshold model. In the fourth section, the multivariate marginal heteroscedasticity model (VAR-A-BEEK) is presented. This model will serve as a comparison method for the leverage effect, and an empiric verification of the theoretical findings will be made with stock market data from Brazil and United States. Finally, the last section presents the conclusions of this paper, and some recommendations for posterior research projects.

## 2. Theoretical Aspects

### 2.1 TAR model

#### 2.1.1 Formal Specifications

The dynamical system characterized by a TAR model consists of an input process $\{Z_t\}$ and an output process $\{X_t\}$, so that

$$X_t = a_0^{(j)} + \sum_{i=1}^{k_j} a_i^{(j)} X_{t-i} + h^{(j)} \varepsilon_t, \; t \in \mathbb{Z}, \tag{1}$$

if $Z_t$ belongs to the real interval $B_j = (r_{j-1}, r_j]$ for any $j = 1, \ldots, l$, with $r_0 = -\infty$ and $r_l = \infty$, so that $\mathbb{R} = \bigcup_{j=1}^{l} B_j$. The coefficients $a_i^{(j)}$ and $h^{(j)}$ ($j = 1, \ldots, l;\; i = 0, 1, \ldots, k_j$) are real numbers and the integer numbers are negative, $k_1, k_2, \ldots, k_l$ make reference to the autoregressive model of order of $\{X_t\}$ in each regime, that is, different autoregressive models of order are permitted in different regimes. The real numbers $r_j$ ($j = 1, \ldots, l-1$) determine the $l$ regimes for the process $\{Z_t\}$ and are named threshold values. Moreover, the process $\{\varepsilon_t\}$ is white Gaussian noise with median 0 and variance 1, mutually independent from $\{Z_t\}$. This model is denoted with the symbol TAR($l; k_1, k_2, \ldots, k_l$).

The TAR($l; k_1, k_2, \ldots, k_l$) model parameters can be divided into two groups:

**i) Structural parameters:** The number of regimes $l$, the $l-1$ threshold $r_1, r_2, \ldots, r_{l-1}$ and autoregressive orders of $l$ regimes $k_1, k_2, \ldots, k_l$.

**ii) Non-structural parameters:** the autoregressive coefficients $a_i^{(j)}$ with $i = 0, 1, \ldots, k_j;\; j = 1, \ldots, l-1$ and the variance weights $h^{(1)}, \ldots, h^{(l)}$.

The threshold autoregressive model presented in (1) is proposed by Tong (1990) in order to specify an open-loop dynamics system; these types of structures are denominated TARSO models (Tong and Lim 1980). The TAR stochastic process can be understood as a nonlinear transfer function, in which its equation shows that the dynamic response of the variable $X_t$ depends on the location of values of the variable $Z_t$ in its sample space (Brockwell and Davis, 1991). The advantages of this type of model are the possibility of obtaining a threshold variable different from the target variable, and the possibility of admitting a wider number of regimens, allowing a bigger generality.

The process $\{Z_t\}$ is considered an exogenous process because there is not feedback of $\{X_t\}$ to it. In the same way, the stochastic behavior of $\{Z_t\}$ is described by a homogenous Markov chain of $m$th order, and as a random variable sequence, converging weakly to the distribution $F$.

### 2.1.2 Conditional and Non-conditional Moments

It is assumed that all the random elements considered in this paper are defined in the same probability space $(\Omega, \mathfrak{F}, P)$. It is possible to compute the marginal distribution $X_t$ using the expression

$$F_t(x) = \sum_{j=1}^{l} p_{t,j} F_{t,j}(x), x \in \mathbb{R}, t \in \mathbb{Z}, \quad (2)$$

with $p_{t,j} = P(Z_t \in B_j)$, $j = 1, ..., l$ (assuming that this probability is positive) and $F_{t,j}(x) = P(X_t \leq x | Z_t \in B_j)$. Consequently, $F_t(x)$ ends up being a mix of accumulated conditional distributions given that $\sum_{j=1}^{l} p_{t,j} = 1$. Hence, given the weak convergence of $\{Z_t\}$ to $F$, considering $p_{t,j} \to p_j = F(r_j) - F(r_{j-1})$ when $t \to \infty$, for all $j = 1, ..., l$ [1].

Under the assumptions of the model (1) and denoting $\mathbb{C}$ as the set of complex numbers, Nieto and Moreno (2016) prove that if the roots of the autoregressive polynomial $\phi_j(z) = 1 - \sum_{i=1}^{k_j} a_i^{(j)} z^i$, $z \in$ $\mathbb{C}$, are located outside the unit circle for each $j = 1, \ldots, l$, the accumulated distribution function for $X_t$ in the regime $j$ is

$$F_{t,j}(x) = \Phi_{0,1}\left(\frac{x - \psi_j(1)a_0^{(j)}}{h^{(j)}\bar{\sigma}_j}\right), x \in \mathbb{R}, \qquad (3)$$

where $\psi_j(z) = \frac{1}{\phi_j(z)} = \sum_{i=0}^{\infty} \psi_i^{(j)} z^i$ for $|z| \leq 1$, with $\sum_{i=0}^{\infty}|\psi_i^{(j)}| < \infty$, $\bar{\sigma}_j^2 = \sum_{i=0}^{\infty}\left(\psi_i^{(j)}\right)^2$ and $\Phi_{0,1}(.)$ expresses the accumulated distribution function of a standard distribution. Under the assumptions considered in (3), it can be demonstrated that $F_{t,j}(x)$ does not depend on $t$, thus $F_{t,j}(x) = F_j(x)$, for all $t$ and for all $j = 1, \ldots, l$.

---

[1] If $\{Z_t\}$ has identical univariate marginal distributions, resembling a stationary process, then the acumulated distribution function of $Z_t$ is $F$ for $t$ therefore $p_{t,j} = p_j$ for $t$ and for $j = 1, ..., l$.

Hence, from this result, and the expected value in (2), it is deduced the expression for the expected non-conditional value

$$E(X_t) = \int_{\mathbb{R}} x \, dF_j(x) = \sum_{j=1}^{l} p_j \int_{\mathbb{R}} x \, dF_j(x), \qquad (4)$$

for all $t$, where $\int_{\mathbb{R}} x \, dF_j(x) = \mu_{j,1}$, defined as the first moment of $X_t$ conditional to the regime $j$, the following is obtained

$$E(X_t) = \sum_{j=1}^{l} p_j \mu_{j,1}. \qquad (5)$$

Then, assuming $\int_{\mathbb{R}} x^2 \, dF_j(x) = \mu_{j,2}$ and knowing that $Var(X_t) = E(X_t^2) - (E(X_t))^2$, the following expression for the non-conditional variance is deduced

$$Var(X_t) = \sum_{j=1}^{l} p_j \mu_{j,2} - \left(\sum_{j=1}^{l} p_j \mu_{j,1}\right)^2, \qquad (6)$$

for all $t$, where $\mu_{j,1} = \frac{a_0^{(j)}}{\phi_j(1)}$, and $\mu_{j,2} = \left(h^{(j)}\bar{\sigma}_j\right)^2 + \left(\frac{a_0^{(j)}}{\phi_j(1)}\right)^2$, with $\phi_j(z) = 1 - \sum_{i=1}^{k_j} a_i^{(j)} z^i$. Through these results, it is verified that the mean functions and variance functions are not conditional, they are constant.

Regarding the conditional moments of the TAR model, there are two types, conditioned by: i) a regimen, $B_j$; ii) a regimen $B_j$ and the set of information until time $t-1$, $\tilde{x}_{t-1} = \{x_{t-1}, \ldots, x_1\}$ with $t > \text{máx}\{k_j | j = 1, \ldots, l\}$; and iii) $\tilde{x}_{t-1}$. Concerning its conditional distributions for i) and ii) normal distribution are obtain, whereas for iii) a mix of conditional normal distributions of i) y ii)[2] are observed (Nieto and Moreno, 2016). Thus, and under the assumption that $\varepsilon_t \sim N(0,1)$, the results presented in the Table 1 are obtained, corresponding to the different means and conditional variances of a TAR model.

**Table 1 here**

Regarding the conditional variance presented on the *Type III* scenario $[Var(X_t|\tilde{x}_{t-1})]$, the extreme value clusters are explained through the regimens of the threshold variable. However, in the ARCH family models, those occur due to the dynamic behavior of the conditional heteroscedasticity.

### 2.1.3 Joint Probability Distribution Function for $X_t$ y $X_{t-\omega}$ and Autocovariance Function

In order to find the autocovariance function for the stochastic process $\{X_t\}$, it is necessary to obtain the accumulated distribution function of the variables $X_t$ y $X_{t-\omega}$ for any integer number $t$ and $\omega$, denoted for $F_{t,t-\omega}$. Under the restrictions of the roots of $\phi_j(z) = 1 - \sum_{i=1}^{k_j} a_i^{(j)} z^i$ in (1) they are located outside the unit circle for each $j = 1, \ldots, l$, then:

$$Cov(X_t, X_{t-\omega}) = \sum_{j,k=1}^{l} p_{t,t-\omega,jk} q_{jk}(\omega) - \mu_t \mu_{t-\omega}, \qquad (7)$$

for all $t, \omega \in \mathbb{Z}$, where $p_{t,t-\omega,jk} = P(Z_t \in B_j, Z_{t-\omega} \in B_k)$ and $q_{jk}(\omega) = \mu_{j,1} \mu_{k,1} + h^{(j)} h^{(k)} \sum_{m=0}^{\infty} \psi_m^{(k)} \psi_{m+\omega}^{(j)}$ for all $j, k = 1, \ldots, l$. In this manner, it is observed that the formal expression for the autocovariance function depends only on the backward $\omega$ and $Cov(X_t, X_{t-\omega}) \to \sum_{j,k=1}^{l} p_{\omega,jk} q_{jk}(\omega) - \mu^2$ when $t \to \infty$. Also, if $\{Z_t\}$ has identical multivariate marginal distributions, it turns out that $Cov(X_t, X_{t-\omega}) = \sum_{j,k=1}^{l} p_{\omega,jk} q_{jk}(\omega) - \mu^2$ (Nieto and Moreno, 2016).

---

[2] Rydén *et al.* (1998) propose modeling the return through a mix of normal variables of zero-mean, demonstrating that many series can be generated with many of the common empiric properties of finances.

It is also important to mention that the TAR stochastic process $\{X_t\}$ presents weak asymptotic stationarity if the conditions given in (3) are satisfied. And it will be a stationary process in a weak-sense, if the stochastic process $\{Z_t\}$ has identical univariate marginal distributions.

### 2.1.4 Conditional Likelihood Function

An important tool of the TAR model is its conditional likelihood function for the structural parameters $l, r_1, \ldots, r_{l-1}, k_1, \ldots, k_l$ and for $\mathbf{x}_k = (x_1, \ldots, x_k)$, where $k = \text{máx}\{k_1, \ldots, k_l\}$, which is given by the joint density function

$$f(\mathbf{y}|\boldsymbol{\theta}_x, \boldsymbol{\theta}_z) = f(\mathbf{x}|\mathbf{z}, \boldsymbol{\theta}_x, \boldsymbol{\theta}_z) f(\mathbf{z}|\boldsymbol{\theta}_x, \boldsymbol{\theta}_z), \qquad (8)$$

with $\mathbf{y} = (\mathbf{x}, \mathbf{z})$, $\mathbf{x}$ and $\mathbf{z}$ data vectors observed for processes $\{X_t\}$ and $\{Z_t\}$, in the sample period $t = 1$ until $t = T$ and $\boldsymbol{\theta}_x$ the vector for all the non-structural parameters. Besides,

$$f(\mathbf{z}|\boldsymbol{\theta}_x, \boldsymbol{\theta}_z) = f(\mathbf{z}_p|\boldsymbol{\theta}_z) f(z_{p+1}|\mathbf{z}_p; \boldsymbol{\theta}_z) \ldots f(z_T|\mathbf{z}_{T-1}; \boldsymbol{\theta}_z), \qquad (9)$$

and

$$f(\mathbf{x}|\mathbf{z}, \boldsymbol{\theta}_x, \boldsymbol{\theta}_z) = f(x_{k+1}|\mathbf{x}_k, \mathbf{z}, \boldsymbol{\theta}_x, \boldsymbol{\theta}_z) \ldots f(x_T|x_{T-1}, \ldots, x_1; \mathbf{z}, \boldsymbol{\theta}_x, \boldsymbol{\theta}_z), \quad (10)$$

where $\mathbf{z}_p = (z_p, \ldots, z_1)$ and $\mathbf{z}_t = (z_t, \ldots, z_{t-p+1})$. Thus, by defining $\{\varepsilon_t\}$ as a white-noise Gaussian process, it is obtained that

$$f(\mathbf{x}|\mathbf{z}, \boldsymbol{\theta}_x, \boldsymbol{\theta}_z) = (2\pi)^{-\frac{(T-k)}{2}} \left[\prod_{t=k+1}^{T} \{h^{(j_t)}\}^{-1}\right] exp\left(-\frac{1}{2}\sum_{t=k+1}^{T} e_t^2\right), \quad (11)$$

with

$$e_t = \frac{x_t - a_0^{(j_t)} - \sum_{i=1}^{k_{j_t}} a_i^{(j_t)} x_{t-i}}{h^{(j_t)}},$$

denoted by $\{j_t\}$ the sequence observed for process $\{J_t\}$, being $\{J_t\}$ a sequence of indicator variables so that $J_t = j$ if and only if $Z_t \in B_j$, for some $j = 1, \ldots, l$. It is necessary to clarify that for the previous equations, it is assumed that there is not an existent relation between the parameters $\boldsymbol{\theta}_x$ and $\boldsymbol{\theta}_z$, and that the density function for $x$ does not depend on $\boldsymbol{\theta}_z$ (Nieto, 2005).

### 2.1.5 Non-linearity Test

As an initial step for adjusting an threshold autoregressive model, it is required to verify the non-linearity of the variable $X_t$. In this paper, the non-linearity test proposed by Nieto and Hoyos (2011) will be used. The authors formulate an extension of the test designed by Tsay (1998).

This methodology is based on the identification of order $k$ of the autoregressive process for the total data of the variable $X_t$ and the selection of the set $S$ of possible values in the lagged parameter $d$, where $d \in \mathbb{Z}^+$. This is with the objective of carrying out a statistic test, in which there is linearity by null hypothesis versus the alternative hypothesis of non-linearity explained with the threshold presence.

### 2.1.6 Identification

For the identification of structural parameters, that is, the number of regimens, the threshold values and the specification of the autoregressive order of each regimen; this research will follow Nieto's proposal (2005). Synthetizing, this strategy is composed by three stages: i) Selecting $l_0$, in order to select the threshold values appropriated for each $l = 2, ..., l_0$, by minimizing the NAIC. In this step, intermediate samples of the non-structural parameters are generated for all the possible combination of autoregressive order; ii) Identifying $l$, based on the intermediate samples of the non-structural parameters and the autoregressive orders; and iii) Identifying the orders $k_1, ..., k_l$, conditional to $l$.

As an advantage of the traditional vision of identification (AIC, graphic methods, etc.), the Bayesian inference permits the construction of a distribution set *a priori* for the number of regimens and the autoregressive orders of each regimen; hence, finding the distributions *a posteriori* (Casella and Robert, 2004; Carlin *et al.*, 2013).

### 2.1.7 Estimation of the non-structural parameters

The conditional density of interest is $p(\boldsymbol{\theta}|\mathbf{x}, \mathbf{z})$, being $\boldsymbol{\theta}$ the vector for unknown parameters of $\{X_t\}$ and $\{Z_t\}$. For obtaining this, the complete conditional densities are calculated for the unknown parameters $\boldsymbol{\theta}_j = \left(a_0^{(j)}, a_1^{(j)}, ..., a_{k_j}^{(j)}\right)'$ ($j = 1, ..., l$), $\mathbf{h} = (h^{(1)}, ..., h^{(l)})'$ and the respective distribution parameters of $\{Z_t\}$. So that $\boldsymbol{\theta}_x = (\boldsymbol{\theta}_1, ..., \boldsymbol{\theta}_l, \mathbf{h})$ and $\boldsymbol{\theta} = (\boldsymbol{\theta}_x, \boldsymbol{\theta}_z)$.

The *a priori* distributions for the component of $\boldsymbol{\theta}_x$, and the densities *a posteriori* of $\boldsymbol{\theta}_j (j = 1, ..., l)$ and $\mathbf{h}$ are presented by Nieto (2005). Once having obtained the complete conditional densities, the Gibbs sampling is used in order to get the estimation of the parameters, taking the statistic samples and the generated samples.

### 2.1.8 Validation of the Model

For each $t = 1, \ldots, T$, be

$$\hat{e}_t = \frac{X_t - X_{t|t-1}}{h^{(j)}},$$

if $Z_t \in B_j$ for some $j$, $j = 1, \ldots, l$, where

$$X_{t|t-1} = a_0^{(j)} + \sum_{i=1}^{k_j} a_i^{(j)} X_{t-i|t-1},$$

is the predictor of $X_t$ one step ahead. The $\hat{e}_t$ are denominated standardized pseudo residuals, and are used in this paper for the verification stage, through the CUSUM and CUSUMSQ graphics. This is in order to check the specification of the model and the heteroscedasticity in $\{\varepsilon_t\}$, respectively. Besides, the simple and partial autocorrelations functions will be used to verify the serial no-correlation in the residuals.

### 3. New mathematic characteristics of the TAR stochastic process

Following, new properties for an open-loop threshold autoregressive model are presented. These properties complement the advances of Nieto (2005, 2008), Moreno (2010), and Nieto and Moreno (2014, 2016). They have been found, specifically, in the third and fourth standardized moment[3] for the interest variable $X_t$, and the mathematical conditions necessary for a specification of the leverage effect are built.

For this section, it is assumed that the marginal univariate distribution of $\{Z_t\}$ are equal, therefore $p_{t,j} = p_j$ for all $t$ and for all $j = 1, \ldots, l$. In this manner, as long as the roots of the polynomial $\phi_j(z) = 1 - \sum_{i=1}^{k_j} a_i^{(j)} z^i$, $z \in \mathbb{C}$, in the model (1) are outside the unit circle, then $\mu = E(X_t)$, which means that the expected value for $X_t$ does not depend on the time.

---

[3] Moreno (2010) tries to find the kurtosis of a TAR model. However, there is a mistake in his calculations.

***Proposition 1.*** Under the conditions described at the beginning of section 3, the analytic expression as a third standardized moment of $X_t$ results as

$$\alpha_3 = \frac{\sum_{j=1}^{l} p_j\left[(\mu_{j,1} - \mu)(3\sigma_j^2 + (\mu_{j,1} - \mu)^2)\right]}{\left[\sum_{j=1}^{l} p_j(\sigma_j^2 + (\mu_{j,1} - \mu)^2)\right]^{\frac{3}{2}}}, \tag{12}$$

*Proof. See Appendix.*

From (12) it is defined that $\alpha_3$ is

$$\begin{cases} < 0, & if \quad \sum_{j=1}^{l} p_j\left[(\mu_{j,1} - \mu)(3\sigma_j^2 + (\mu_{j,1} - \mu)^2)\right] < 0, \\ \geq 0, & if \quad \sum_{j=1}^{l} p_j\left[(\mu_{j,1} - \mu)(3\sigma_j^2 + (\mu_{j,1} - \mu)^2)\right] \geq 0. \end{cases}$$

As a main result of this proposition, it is found that the sign of the coefficient of asymmetry ends up depending essentially on the difference between expected value of $X_t$ conditioned to the regimen $j$, $E(X_t | Z_t \in B_j) = \psi_j(1) a_0^{(j)}$, and the marginal expected value of $X_t$, $E(X_t) = \sum_{j=1}^{l} p_j \mu_{j,1}$.

***Proposition 2.*** Assuming that the univariate marginal distributions of $\{Z_t\}$ are equal, the analytic expression of the kurtosis as a fourth standardized moment of $X_t$ turns out to be

$$\alpha_4 = \frac{\sum_{j=1}^{l} p_j\left[(\mu_{j,1} - \mu)^4 + 6\sigma_j^2(\mu_{j,1} - \mu)^2 + 3\sigma_j^4\right]}{\left[\sum_{j=1}^{l} p_j(\sigma_j^2 + (\mu_{j,1} - \mu)^2)\right]^2}. \tag{13}$$

*Proof. See Appendix*

It is important to highlight that it is necessary to be careful with the interpretation of the kurtosis in the case of a TAR model, because the distribution of the variable $X_t$,

$$F(x) = \sum_{j=1}^{l} p_j F_j(x), x \in \mathbb{R}, j = 1, \ldots, l,$$

is a mix of accumulated conditional distributions given that $\sum_{j=1}^{l} p_j = 1$ and consequently $F(x)$ is multimodal.

In the international literature there is still not clarity about the kurtosis coefficient interpretation of a multimodal distribution (Knapp, 2007; Cahoy, 2015; Chakraborty, Hazarika and Ali, 2015). Nevertheless, authors such as Finucan (1964), Darlington (1970) and DeCarlo (1997) affirm that having a distribution with different peaks, it would be difficult to get big excesses of kurtosis.

***Empirical proof by simulation.*** Three simulations of weak stationary autoregressive stochastic processes of threshold are carried out, in order to calculate its asymmetry and sample kurtosis, and compare them with the results of the equations (12) and (13).

**Model 1 (M1).** The process TAR(2;0,1) is considered, given by

$$X_t = \begin{cases} 0.6 + 0.7\varepsilon_t, & if\ Z_t \leq 0, \\ 0.2 + 0.4X_{t-1} + 1.1\varepsilon_t, & if\ Z_t > 0, \end{cases} \quad (14)$$

where $\varepsilon_t \sim RBG(0,1)$, $Z_t = 0.5Z_{t-1} + \tau_t$ with $\tau_t \sim RBG(0,1)$. Here $p_1 = p_2$.

**Model 2 (M2).** The process TAR(2;2,3) is considered, given by

$$X_t = \begin{cases} 2.9 + 0.3X_{t-1} - 0.4X_{t-2} + 1.5\varepsilon_t, & if\ Z_t \leq 0, \\ 0.6 - 0.3X_{t-1} - 0.1X_{t-2} + 0.2X_{t-3} + \varepsilon_t, & if\ Z_t > 0 \end{cases} \quad (15),$$

where $\varepsilon_t \sim RBG(0,1)$, $Z_t = 0.4Z_{t-1} + \tau_t$ with $\tau_t \sim RBG(0,0.5)$. It is obtained that $p_1 = p_2$.

**Model 3 (M3).** The process TAR(3;3,1,3) is considered, given by

$$X_t = \begin{cases} -1.6 + 0.2X_{t-1} - 0.6X_{t-2} - 0.1X_{t-3} + 3\varepsilon_t, & if\ Z_t \leq -1, \\ 0.9 + 0.5X_{t-1} + \varepsilon_t, & if\ -1 < Z_t \leq 1, \\ 4 - 0.7X_{t-1} - 0.2X_{t-2} + 0.1X_{t-3} + 2\varepsilon_t, & if\ Z_t > 1, \end{cases} \quad (16),$$

where $\varepsilon_t \sim RBG(0,1)$, $Z_t = 0.6Z_{t-1} + \tau_t$ with $\tau_t \sim RBG(0,1)$. In this model $p_2 = 0.6$ y $p_1 = p_3$.

Each model (M1, M2, y M3) was simulated 1000 times with a sample size of 300. In each exercise time series of 600 lengths were simulated, dismissing the first 300 for decreasing the effect of input values. Then, the asymmetry and the kurtosis of each of the 1000 series was computed, and their mean was calculated, calling these numbers the mean of the sample asymmetry $(\hat{\alpha}_3)$ and the mean of the sample kurtosis $(\hat{\alpha}_4)$. Additionally, the standard deviations of these 1000 time series were $(ds_{\hat{\alpha}_3}, ds_{\hat{\alpha}_4})$ and based on this, the intervals $(\hat{\alpha}_3 \pm 2ds_{\hat{\alpha}_3})$ and $(\hat{\alpha}_4 \pm 2ds_{\hat{\alpha}_4})$ were constructed, to determine if the theoretical coefficient values of the asymmetry $(\alpha_3)$ and the kurtosis $(\alpha_4)$ were in their respective sample intervals.

In the Table 2, there are shown data concerning the analysis of the sample asymmetry for M1, M2 and M3. It is observed that the theoretical asymmetry coefficient for the three models $(\alpha_3)$, calculated through the equation (12), has the same sign that the mean of the sample asymmetry $(\hat{\alpha}_3)$ and it is found within the interval $(\hat{\alpha}_3 \pm 2ds_{\hat{\alpha}_3})$. Thus, it is demonstrated that the theoretical values are very close to the sample-estimated values.

**Table 2 here**

In the Table 3 there are exposed the results regarding the sample calculations of the kurtosis for the same three models of analysis. For all of the cases the kurtosis coefficient is included in the interval $(\hat{\alpha}_4 \pm 2ds_{\hat{\alpha}_4})$.

**Table 3 here**

*Analytic specification of the conditional variance Type III to represent the leverage effect in the TAR model.* Having studied the mathematical structure of an autoregressive model of thresholds, and knowing, by previous works, that it has been demonstrated that it can represent certain characteristics of the financial series; the current research states that an approximation to the leverage effect is possible, under certain conditions of the conditional variance of the past data of the interest variable.

The accumulated distribution function of $X_t$ given $\tilde{x}_{t-1}$ is defined by (Nieto and Moreno, 2016)

$$F_t(X_t|\tilde{x}_{t-1}) = \sum_{j=1}^{l} p_j P\left((X_t \leq x)\big|Z_t \in B_j, \tilde{x}_{t-1}\right), \qquad (17)$$

from which

$$\begin{aligned} E(X_t|\tilde{x}_{t-1}) &= \sum_{j=1}^{l} p_j \left( a_0^{(j)} + \sum_{i=1}^{k_j} a_i^{(j)} x_{t-i} \right) \\ &= \sum_{j=1}^{l} p_j a_0^{(j)} + \sum_{j=1}^{l} p_j a_1^{(j)} x_{t-1} + \cdots + \sum_{j=1}^{l} p_j a_{k_j}^{(j)} x_{t-k_j}, \end{aligned} \qquad (18)$$

and

$$\begin{aligned}
Var(X_t|\tilde{x}_{t-1}) &= \sum_{j=1}^{l} p_j \left( (h^{(j)})^2 + \left( a_0^{(j)} + \sum_{i=1}^{k_j} a_i^{(j)} x_{t-i} \right)^2 \right) \\
&\quad - \left( \sum_{j=1}^{l} p_j \left( a_0^{(j)} + \sum_{i=1}^{k_j} a_i^{(j)} x_{t-i} \right) \right)^2 \\
&= \sum_{j=1}^{l} p_j (h^{(j)})^2 + \sum_{j=1}^{l} p_j \left( a_0^{(j)} + \sum_{i=1}^{k_j} a_i^{(j)} x_{t-i} \right)^2 \\
&\quad - \left( \sum_{j=1}^{l} p_j \left( a_0^{(j)} + \sum_{i=1}^{k_j} a_i^{(j)} x_{t-i} \right) \right)^2.
\end{aligned} \quad (19)$$

From (19), it is observed that having the fixed parameters $p_j, h^{(j)}, a_0^{(j)}$ and $a_i^{(j)}, i = 1, \ldots, k_j$, the conditional variance of the model ends up in function of the values that $x_{t-i}$ takes, affecting directly its volatility change, $\sqrt[2]{Var(X_t|\tilde{x}_{t-1})}$, similar event to what occurred with the variation effect on the innovations $a_{t-i}$ under the conditional standard deviation of a model from the ARCH family.

Knowing that $a_{t-i}$ and $x_{t-i}$ are different concepts, it could represent certain leverage effect through the information given by (19), assuming that the realizations of the variable of interest incorporate part of the information of the innovations of the model. Resuming with the technical definition of the *leverage effect*, which occurs when the volatility of a falling stock market is superior to the volatility of rising stock market. Comparing the square root of $Var(X_t|\tilde{x}_{t-1})$ versus $x_{t-i}$, turns out to be a similar, but not identical, approach to that built based on the News Impact Curves (NIC).

Therefore, assuming that the square root of $Var(X_t|\tilde{x}_{t-1})$ is defined as the type of volatility that better represents the characterization of financial series from an open-loop TAR model, it is feasible to calculate how the variations in $x_{t-i}$ affect its volatility. Having said that, supposing by interpretability that in the last returns the variable was of the same magnitude $x^*$, then

$$\sqrt[2]{[Var(X_t|\tilde{x}_{t-1})]}$$

$$= \sqrt[2]{\left[\sum_{j=1}^{l} p_j(h^{(j)})^2 + \sum_{j=1}^{l} p_j\left(a_0^{(j)} + \sum_{i=1}^{k_j} a_i^{(j)} x^*\right)^2 - \left(\sum_{j=1}^{l} p_j\left(a_0^{(j)} + \sum_{i=1}^{k_j} a_i^{(j)} x^*\right)\right)^2\right]}. \quad (20)$$

By analyzing the functional form of (20), it is similar to what is commonly represented in the NIC, where, by studying the impact of the innovations in the previous period over the present volatility, ends up creating graphically a concave curve upwards. In order to proof mathematically this, in the expressions (21), (22) and (23) are presented the first derivative respect to $x^*$, its only root and the second derivative respect to $x^*$, respectively.

$$\frac{\partial Var(X_t|\tilde{x}_{t-1})}{\partial x^*}$$

$$= \sum_{j=1}^{l}\left[p_j\left(2a_0^{(j)} \sum_{i=1}^{k_j} a_i^{(j)}\right)\right] + \sum_{j=1}^{l}\left[p_j\left(\sum_{i=1}^{k_j} a_i^{(j)}\right)^2\right] 2x^* \quad (21)$$

$$- 2\sum_{j=1}^{l}\left[p_j a_0^{(j)}\right] \sum_{j=1}^{l}\left[p_j \sum_{i=1}^{k_j} a_i^{(j)}\right] - \left(\sum_{j=1}^{l}\left[p_j \sum_{i=1}^{k_j} a_i^{(j)}\right]\right)^2 2x^*,$$

$$x^* = \frac{\sum_{j=1}^{l}\left[p_j a_0^{(j)}\right] \sum_{j=1}^{l}\left[p_j \sum_{i=1}^{k_j} a_i^{(j)}\right] - \sum_{j=1}^{l}\left[p_j\left(a_0^{(j)} \sum_{i=1}^{k_j} a_i^{(j)}\right)\right]}{\sum_{j=1}^{l}\left[p_j \left(\sum_{i=1}^{k_j} a_i^{(j)}\right)^2\right] - \left(\sum_{j=1}^{l}\left[p_j \sum_{i=1}^{k_j} a_i^{(j)}\right]\right)^2}, \quad (22)$$

$$\frac{\partial^2 Var(X_t|\tilde{x}_{t-1})}{\partial x^{*2}} = 2\left[\sum_{j=1}^{l}p_j\left(\sum_{i=1}^{k_j}a_i^{(j)}\right)^2 - \left(\sum_{j=1}^{l}p_j\sum_{i=1}^{k_j}a_i^{(j)}\right)^2\right]$$

$$= 2\left[p_1(a_1^1+\cdots+a_{k_1}^1)^2+\cdots+p_l(a_1^l+\cdots+a_{k_l}^l)^2\right.$$

$$-p_1^2(a_1^1+\cdots+a_{k_1}^1)^2-\cdots-p_l^2(a_1^l+\cdots+a_{k_l}^l)^2 \qquad (23)$$

$$-2p_1p_2(a_1^1+\cdots+a_{k_1}^1)(a_1^2+\cdots+a_{k_2}^2)-\cdots$$

$$\left.-2p_{l-1}p_l(a_1^{l-1}+\cdots+a_{k_{l-1}}^{l-1})(a_1^l+\cdots+a_{k_l}^l)\right].$$

By observing that the second derivative is not in function of de $x^*$, its sign can be analyzed to determine the concavity without using the calculations of (22). Under the following sufficient condition

$$\sum_{i=1}^{l}\sum_{j=1}^{l}p_ip_j\left(\sum_{m=1}^{k_i}a_m^{(i)}\right)\left(\sum_{n=1}^{k_j}a_n^{(j)}\right) \leq 0, \qquad (24)$$

is ensured a second derivative (23) strictly positive, from which is obtained that the function (20) in the point $x^*$ is concave upwards, and being this the only critical point, the function (20) is concave upwards in all its domain.

Then, it is important to highlight that the expression (22) turns out to be the value of $x^*$ where the volatility is minimum for the $TAR(l;k_1,\ldots,k_l)$ model. This value is denoted $x_{mín}^*$.

Thus, it is important to underline that if $x_{mín}^*$ is strictly greater than zero, $x_{mín}^* > 0$, it manages to capture the concept of leverage effect, under this condition

$$f(x^*) = \sqrt[2]{[Var(X_t|\tilde{x}_{t-1}=(x^*,\ldots,x^*))]} < f(-x^*).$$

This means that the realization of negative return will affect the volatility of the financial product in a greater measurement than the realization of return with the same magnitude but with a positive sign.

## 4. An application with stock market series

This section evaluates the adjustment of the TAR model proposed by Nieto (2005), in the analysis of financial series compared with a MGARCH system of type A-BEEK, using the conditional and non-conditional moments and the representation of the leverage effect.

The Bovespa stock market index is modeled, specifically, as a variable of interest, and S&P 500 index as a threshold variable. The demand capacity of United States, the influence of the Federal Reserve over the World's Central Banks, and the relevance of the companies that pay contributions in this country, among others; are reasons enough to know that the S&P 500 has a supported relation with the interest index. The empirical support for Brazil is adduced by Ozun and Ozbakis (2007).

The information of the stock market indexes is extracted form the platform *Thomson Reuters Eikon*, taking the datum of the daily closure of the business days for both countries, following the proposals of Connolly (1989), Susmel and Engle (1994), and Pérez and Torra (1995). For the adjustment of the models, the missing data[4] are estimated, according to the methodology of Nieto and Ruiz (2002)[5]. In this way, complete information is available from Monday to Friday for the different series.

The data correspond to the observation registered from January 2 of 2009 to June 23 of 2015, obtaining a total of 1688 data. This time period contemplates a new architectonical structure of the international financial system, after the *subprime* crisis originated in 2007. From the beginning of the contemplated time interval, a new legislation was erected, focused on controlling the systemic risk, and augmenting the stock market agents trust (Kodres and Narain, 2012).

---

[4] These make reference to holydays or days in which the Stock Markets were not operating.
[5] By estimating the missing values, it results relevant because homogeneous financial series are obtained, and the loss of fundamental information can be avoided. Besides, the methodology used for estimating the missing data is a reasonable statistics process.

### 4.1  Return of the Bovespa and S&P 500 stock market indexes

For this application, the financial returns[6] are defined in this way:

$$X_t = ln(Bovespa_t) - ln(Bovespa_{t-1})$$

$$Z_t = ln(S\&P\ 500_t) - ln(S\&P\ 500_{t-1}).$$

In this manner, the Figure 1 illustrates the dynamic of the Bovespa and S&P 500 stock markets indexes, and their calculated returns, in which certain conglomerate of external values[7] are observed.

**Figure 1 here**

In the Figure 2, there is a comparison between the empiric distributions of the Bovespa return series, and a simulated normal distribution that assumes the same standard deviation of the series (0.0145), confirming the presence of long and heavy queues (higher probability of external events) in the empiric distribution of the Bovespa returns.

**Figure 2 here**

---

[6] Such return series (also denominated as logarithmic returns series), is the most used measure in financial modeling because it is stable in measure, does not contemplate units and simplify the calculation of the composed return $k$ periods between the time $t - k$ and $t$ (Fan and Yao, 2003; Tsay, 2014).
[7] Applying the unit root Augmented Dickey-Fuller test (ADF), Phillip-Perron (PP) and Kwiatkowski-Phillips-Schmidt-Shin (KPSS), to a significance level of 5%, there is statistical evidence that shows that both returns are stationary.

### 4.2 Adjustment of the TAR model to the Bovespa returns series

Using the adjustment methodology of section 2, which follows Nieto's approximation (2005), it was found the following TAR model for the Bovespa financial returns:

$$X_t = \begin{cases} -0.0059 + 0.0132\varepsilon_t & \text{If } Z_t \leq 0.0007, \\ 0.0063 - 0.0478X_{t-1} - 0.0533X_{t-2} - 0.0779X_{t-3} & \text{If } Z_t > \\ -0.0171X_{t-4} - 0.0508X_{t-5} - 0.0653X_{t-6} + 0.0128\varepsilon_t & 0.0007. \end{cases} \quad (25)$$

In the Table 4, it is represented the estimation of the parameters, the value of the square root of the function of the mean squared loss calculated in the Bayes optimal estimation, which, from this moment, will be referred simply as typical deviation, and a 90% credible interval. The results express the statistical significance level 10% of almost all the coefficients, and minor differences between the variances *Type II* of both regimens. By performing a sensitivity analysis of such estimations, it is observed that the coefficients do not perceive major alterations when the values *a priori* are changed.

**Table 4 here**

Regarding the model validation, the functions of simple and partial autocorrelation of the standardized residuals indicate the serial no correlation (Figure 3 (a) and (b)). The CUSUM and CUSUMSQ graphics (Figure 4 (a) and (b)) do not evidence a bad model specification nor a marginal heteroscedasticity in the variable $\{\varepsilon_t\}$, with a 95% of confidence level. Therefore, according to the related literature, the model presents a good fit.

**Figure 3 here**

**Figure 4 here**

## 4.3 Adjustment of the asymmetric multivariate GARCH model to the Bovespa return series

### 4.3.1 VAR-A-BEKK Model

Particularly, for MGARCH models that manage to capture the leverage effect, there can be found approaches such as VARMA-AGARCH of McAleer, Hoti and Chan (2009), asymmetric dynamic covariance (generalization that includes the asymmetric versions of the VECH, CCC, BEKK and FARCH) of Kroner and Ng (1998), the GARCH matrix exponential of Kawakatsu (2006) and the GARCH with dynamic asymmetry -DAMGARCH- of Caporin and McAleer (2011).

For this research, it was decided to use the asymmetric BEKK model, denoted A-BEKK, as a comparison referent for the TAR model. Having said that, in order to model the correlation of returns and their mean $f(t-1)$, a Vector Autoregressive system (VAR) is chosen. Thus, a multivariate VAR(p)-A-BEKK(1,1) model is adopted. Formally,

$$\mathbf{R_t} = \mathbf{\mu} + \sum_{j=1}^{p} \mathbf{\Gamma_j} \mathbf{R_{t-j}} + \mathbf{a_t} \qquad (26)$$

where $\mathbf{R_t}$, $\mathbf{\mu}$, $\mathbf{R_{t-j}}$ and $\mathbf{a_t}$ are vectors Nx1, with $t = 1, \ldots, T$ being $\{1, \ldots, T\}$ the period of observation of the series; $\mathbf{\Gamma_j}$ has dimension NxN, with $j = 1, \ldots, p$. Moreover, $\mathbf{a_t}|F_{t-1} \sim N_N(\mathbf{0}, \mathbf{H_t})$. Under these assumptions, the conditional mean of the model is described by a process VAR(p),

$$E[\mathbf{R}_t|\mathbf{F}_{t-1}] = \mu + \sum_{j=1}^{p} \Gamma_j \mathbf{R}_{t-j}. \tag{27}$$

The model assumes that the series $\{\mathbf{R}_t; t = 1, \dots, T\}$ behaves conditionally heteroscedastic with matrix variances and covariances $H_t$, defined positive[8], $Var[\mathbf{R}_t|\mathbf{F}_{t-1}] = H_t$, Which evolution follows a process A-BEKK(1,1) given by

$$\begin{aligned} H_t = C'C &+ \lambda' \mathbf{a}_{t-1} \mathbf{a}_{t-1}' \lambda + \vartheta' H_{t-1} \vartheta \\ &+ D'\big(\mathbf{a}_{t-1} \circ I(\mathbf{a}_{t-1} < 0)\big)\big(\mathbf{a}_{t-1} \circ I(\mathbf{a}_{t-1} < 0)\big)' D, \end{aligned} \tag{28}$$

where $\mathbf{a}_t$ is a matrix Nx1, $C$ is a matrix NxN upper triangular, $\lambda$, $D$ and $\vartheta$ are matrix parameters NxN and $\circ$ denotes the vector product component by component. The expression $\lambda' \mathbf{a}_{t-1} \mathbf{a}_{t-1}' \lambda$ is denominated ARCH effect, and represents the impact exerted by the past innovation $\mathbf{a}_{t-1}$. On the other hand, $\vartheta' H_{t-1} \vartheta$ is known as the GARCH effect, and describes the impact exerted by the conditional volatility of the process in the previous period.

The last term of (28), $D'\big(\mathbf{a}_{t-1} \circ I(\mathbf{a}_{t-1} < 0)\big)\big(\mathbf{a}_{t-1} \circ I(\mathbf{a}_{t-1} < 0)\big)' D$, permits to capture the asymmetric effects, contemplating the leverage effect, over the components of the matrix $H_t$. The inclusion of this expression permits to capture different relative responses to positive and negative *shocks* in the matrix of variances and covariances.

The parameters of the VAR(p)-A-BEKK(1,1) model are estimated with the maximum likelihood estimation method. Then, the logarithm of the function of likelihood is given by

$$l(\boldsymbol{\theta}) = \sum_{t=1}^{T} l_t(\boldsymbol{\theta}), \tag{29}$$

---

[8] $H_t$ is a matrix defined positive under the sufficient condition that at least one of the matrices $C$ or $\vartheta$ have full rank (Engle and Kroner, 1995).

where $l_t(\boldsymbol{\theta}) = -\ln(2\pi) - \frac{1}{2}\ln(|\mathbf{H_t}(\boldsymbol{\theta})|) - \frac{1}{2}\mathbf{a_t}'(\boldsymbol{\theta})\mathbf{H_t^{-1}}(\boldsymbol{\theta})\mathbf{a_t}(\boldsymbol{\theta})$, with $\boldsymbol{\theta}$ the parameter vector of the model that includes all the non-null components of the vector $\boldsymbol{\theta}$ and the matrices $\{\boldsymbol{\Gamma_j}; j = 1, ..., p\}$, $\boldsymbol{C}$, $\boldsymbol{\lambda}$, $\boldsymbol{D}$ and $\boldsymbol{\vartheta}$.

It is important to highlight that the A-BEKK approximation was selected due to the following arguments:

i) By estimating different types of MGARCH (A-VEC, A-CCC, A-DCC, among others), it is the one that gets the best results via criteria minimizing, Akaike (*AIC*) and Bayesian (*BIC*);

ii) It ensures $H_t$ positive-definite matrices, due to the quadratic nature of its equations. Moreover, it is more general than the diagonal representations, because it allows certain relations that the latters do not admit, such as the fact that the lagged variance affect the dynamic of the covariances (Franses and van Dijk, 2000);

iii) In the international literature the A-BEKK present one of the best adjustments for empiric application when MGARCH are estimated with leverage effect (Isakov and Pérignon, 2001; Brooks, Henry and Persand, 2002; Tai, 2004; Henry, Olekalns and Shields, 2010; Asai and McAleer, 2011; Rahman and Serletis, 2012, among others);

iv) In contrast with other restrictive models, such representation permits a wider dynamic, by not assuming the matrix of the conditional correlations as constant (Tsay, 2014).

### 4.3.2 Leverage effect representation in the VAR-A-BEKK model

Within the univariate time series field, Pagan and Schwert (1990), and Engle and Ng (1993) came up with the construction of the denominated News Impact Curve (NIC), where it is examined the relation between different changes in the lagged innovation $a_{i,t-1}$ and its consequence in the conditional volatility $\sigma_{i,j,t}$, with the information until the period $t-2$ constant, denoted $\boldsymbol{F_{t-2}}$. Kroner and Ng (1998) extend the NIC to the case of multivariate time series, with the denominated News Impact Surface (NIS)[9]. Such generalization estimates the conditional variances and covariances against *shocks* of the immediately previous period in the returns, maintaining the constant past variances-covariances.

For the bivariate A-BEKK(1,1) model, the equations in the NIS are described by the following analytic expressions:

$$\sigma_{1,1,t} = \hat{c}_{1,1}^2 + \hat{\lambda}_{1,1}^2 a_{1,t-1}^2 + 2\hat{\lambda}_{1,1}\hat{\lambda}_{2,1} a_{1,t-1} a_{2,t-1} + \hat{\lambda}_{2,1}^2 a_{2,t-1}^2 + \hat{\vartheta}_{1,1}^2 \overline{\sigma_{1,1,t-1}^2}$$
$$+ 2\hat{\vartheta}_{1,1}\hat{\vartheta}_{2,1} \overline{\sigma_{2,1,t-1}^2} + \hat{\vartheta}_{2,1}^2 \overline{\sigma_{2,2,t-1}^2} + \hat{d}_{1,1}^2 \zeta_{1,t-1}^2 + 2\hat{d}_{1,1}\hat{d}_{2,1} \zeta_{1,t-1}\zeta_{2,t-1} \quad (30)$$
$$+ \hat{d}_{2,1}^2 \zeta_{2,t-1}^2.$$

$$\sigma_{2,2,t} = \hat{c}_{2,2}^2 + \hat{c}_{1,2}^2 + \hat{\lambda}_{1,2}^2 a_{1,t-1}^2 + 2\hat{\lambda}_{1,2}\hat{\lambda}_{2,2} a_{1,t-1} a_{2,t-1} + \hat{\lambda}_{2,2}^2 a_{2,t-1}^2 + \hat{\vartheta}_{1,2}^2 \overline{\sigma_{1,1,t-1}^2}$$
$$+ 2\hat{\vartheta}_{1,2}\hat{\vartheta}_{2,2} \overline{\sigma_{2,1,t-1}^2} + \hat{\vartheta}_{2,2}^2 \overline{\sigma_{2,2,t-1}^2} + \hat{d}_{1,2}^2 \zeta_{1,t-1}^2 + 2\hat{d}_{1,2}\hat{d}_{2,2} \zeta_{1,t-1}\zeta_{2,t-1} \quad (31)$$
$$+ \hat{d}_{2,2}^2 \zeta_{2,t-1}^2.$$

$$\sigma_{1,2,t} = \sigma_{2,1,t} = \hat{c}_{1,1}\hat{c}_{1,2} + \hat{\lambda}_{1,1}\hat{\lambda}_{2,2} a_{1,t-1}^2 + (\hat{\lambda}_{2,1}\hat{\lambda}_{1,2} + \hat{\lambda}_{1,1}\hat{\lambda}_{2,2}) a_{1,t-1} a_{2,t-1}$$
$$+ \hat{\lambda}_{2,1}\hat{\lambda}_{2,2} a_{2,t-1}^2 + \hat{\vartheta}_{1,1}\hat{\vartheta}_{1,2} \overline{\sigma_{1,1,t-1}^2} + (\hat{\lambda}_{2,1}\hat{\lambda}_{1,2} + \hat{\lambda}_{1,1}\hat{\lambda}_{2,2}) \overline{\sigma_{1,2,t-1}^2}$$
$$+ \hat{\vartheta}_{2,1}\hat{\vartheta}_{2,2} \overline{\sigma_{2,2,t-1}^2} + \hat{d}_{1,1}\hat{d}_{2,2} \zeta_{1,t-1}^2 + (\hat{d}_{2,1}\hat{d}_{1,2} + \hat{d}_{1,1}\hat{d}_{2,2}) \zeta_{1,t-1}\zeta_{2,t-1} \quad (32)$$
$$+ \hat{d}_{2,1}\hat{d}_{2,2} \zeta_{2,t-1}^2,$$

where $\hat{c}, \hat{\lambda}, \hat{\vartheta}$ and $\hat{d}$ are the estimators of the different coefficients of the MGARCH model. Additionally, given that the conditional variances and covariances $\sigma_{i,j,t}^2$ in the A-BEKK model depend on their own past, it is reasonable to fix the past conditional variances and covariances in their conditional mean values throughout the sample period, detonating them as $\overline{\sigma^2_{1,1,t-1}}, \overline{\sigma^2_{2,2,t-1}}$ y $\overline{\sigma^2_{1,2,t-1}}$. Then, through the equations (30), (31) and (32), it will be possible to observe the variations of $\sigma^2_{1,1,t}$, $\sigma^2_{2,2,t}$, $\sigma^2_{1,2,t}$ and $\sigma^2_{2,1,t}$, facing changes in $a_{1,t-1}$ and $a_{2,t-1}$.

---

[9] The articles by Jondeau and Rockinger (2009), and Caporin and McAleer (2011) share an excellent theoretical reference for understanding the NIS philosophy.

Another perspective to be taken into account, in order to observe the asymmetry in the volatility, and therefore, the leverage effect; is the approximation proposed in the seminal paper of French, Schwert and Stambaugh (1987), in which it is proposed the construction of a return between the percent change of the conditional standard deviation estimated from the active return $i$, $\sigma_{i,t}$, and the return of the same financial product, $r_{i,t}$,

$$ln\left(\frac{\sigma_{i,t}}{\sigma_{i,t-1}}\right) = \alpha_0 + \alpha_1 r_{i,t-1} + \varepsilon_t, \qquad (33)$$

where by denoting $P_t$, with $t \in \mathbb{Z}$, as the price series of a financial asset $t$, and knowing that the simple returns $R_t$ are defined as $\frac{P_t - P_{t-1}}{P_{t-1}}$, the logarithmic returns can be defined as

$$r_t = ln(1 + R_t) = ln\left(1 + \frac{P_t}{P_{t-1}} - 1\right) = ln(P_t) - ln(P_{t-1}),$$

expressing how different is the price between one time instant and other in logarithmic scale[10] (Campbell, Lo and MacKinlay, 1997).

In order to represent the *leverage effect*, Black (1976) and Christie (1982) stated that such estimation should have a negative elasticity ($\alpha_1 < 0$). Besides, they affirmed that it the elasticity is higher than a -1, this could indicate that the leverage effect would be the only reason of the negative relation between the returns and the volatility of the reference asset.

---

[10] If it is assumed that the prices have a log-normal distribution, then $ln(1 + R_t)$ it is normal distributed. Thus, using this type of expression permits the characteristic of time additivity, given that the addition of non-correalted normal variables returns to be a normal variable.

### 4.3.3 Empiric application of the VAR-A-BEKK model

The first measure is to carry out the effect tests ARCH of Engle (1982) over a bivariate VAR(1) model, chosen taking into account the criteria of Schwarz (SIC) and the final prediction error (FPE), adjusted to the Bovespa and S/P500 stock market indexes[11], in which it was evidenced, through the statistics of the Lagrange multiplier, that for any conventional significance level the null hypothesis of no effects ARCH is rejected. This allowed supporting the pertinence of a conditional heteroscedasticity model.

Therefore, a VAR(1)-BEKK(1,1) model was estimated in order to evaluate the diagnosis of the asymmetry in the volatility. The statistic joint test of Engle and Ng (1993) was applied, such test has as its null hypothesis the no existence of asymmetries (Brooks, 2009), and the leverage effect test of Enders (2014), which has as its null hypothesis the no existence of such effects[12]. For both series, the absence of asymmetries is rejected to a 10% significance level, justifying in this way the estimation of a model that considers asymmetric effects such as VAR(1)-A-BEKK(1,1).

The estimated coefficients of the VAR(1)-A-BEKK(1,1) model are shown in the Table 5, along with the joint validation test over its residuals. The results for this model indicate that the majority of elements of the matrices $\mu$, $\Gamma$, $C$, $\lambda$, $\vartheta$ and $D$ are significant at 10%. Besides, the null hypotheses of non-autocorrelation and marginal heteroscedasticity are not rejected. On the other hand, the normality test of the standardized residuals evidences the persistence of kurtosis excess, which entails the rejection of the null hypothesis of normality.

---

[11] The modules of the inverse roots of the polynomials associated with the model VAR(1) were 0.0894 and 0.0336, demonstrating that the process posseses the stability condition (stationarity), because they are values smaller than the unit (Lütkepohl, 2005).

[12] This, carries out a global significance test based on the following regression, $\eta_t^2 = \delta_0 + \delta_1 \eta_{t-1} + \delta_2 \eta_{t-2} + \cdots + \delta_k \eta_{t-k} + v_t$, establishing itslef as the null hypothesis $H_0: \delta_1 = \cdots = \delta_k = 0$ (there are not leverage effects), and alternative hypothesis, $H_1$: at least a $\delta_i \neq 0$ (there are leverage effects) (Enders, 2014).

**Table 5**

## 4.4 Calculation and comparison of the conditional and non-conditional moments of the TAR and MGARCH models for the Bovespa return series

For the TAR model presented in the equation (25) we have that $p_1 = 0.5$ and $p_2 = 0.5$ [13]. In the same way, when reviewing the roots of the characteristic polynomials on each regimen, it was verified that they are found outside the unit circle. By knowing that, it was possible to find coefficients of the inverse polynomials given by $\psi^{(j)}(\mathcal{Y}) = 1 - \psi_1^{(j)}\mathcal{Y} - \psi_2^{(j)}\mathcal{Y}^2 - \psi_3^{(j)}\mathcal{Y}^3 - \cdots$. Through this information it is obtained that $\psi^{(1)}(1) = 1$, $\psi^{(2)}(1) = 0.7621$, $\bar{\sigma}_1^2 = 1$ y $\bar{\sigma}_2^2 = 1.015$. Also, the values $\mu_{t,j,1} = \frac{a_0^{(j)}}{\phi_j(1)}$ and $\mu_{t,j,2} = \left(h^{(j)}\sigma_j\right)^2 + \left(\frac{a_0^{(j)}}{\phi_j(1)}\right)^2$ $(j = 1,2)$, fundamentals for the calculation of the first four non-conditional moments are presented in the Table 6.

**Table 6 here**

Moreover, in the Table 7 are shown the conditional and non-conditional moments of the model TAR(2;0,6) estimated for the Bovespa return series. The TAR model for the Brazilian stock market index contemplates two regimens with characteristics clearly differentiated. The first regimen regards negative returns (and some positive close to zero) in the North American stock market; while the second regimen regards only positive returns of the S&P 500.

---

[13] In this application, $p_{t,j}$ does not depend on $t$, given that it is assumed that the marginal distributions of the financial series are identical for all $t$ and for all $j = 1, \ldots, l$.

**Table 7 here**

As it was expected, in the first regimen it was obtained a loss of -0.59% and a conditional standard deviation *Type I* of the 1.41%. For the second regimen, the conditional mean *Type I* is of 0.48% with a deviation equal to that in the first regimen. It is observed that the conditional mean *Type II* is of -0.59% for the first regimen, while for the second regimen it depends on the returns of the previous six days.

Besides, the conditional standard deviation *Type II* ends up being the same of that of *Type I* (1.41%). On the other hand, the daily marginal return expected for the Bovespa returns is of -0.05% with a standard deviation of 1.41%. The asymmetry turns out to be negative and the kurtosis slightly minor to three, reflecting a platykurtic distribution.

In the Table 8 are shown the conditional and non-conditional moments of the MGARCH model for the Bovespa return series, where it is observed an expected return of 0.02% and a marginal standard deviation equal to that of the TAR model (1.41%). This stochastic process also contemplates a negative asymmetry, but a kurtosis higher than 3.

**Table 8 here**

By observation the results of the TAR and MGARCH models, it is possible to highlight that both adjust, relatively good, the behavior of the Brazilian returns.

Additionally, in Table 9, it is deduced that for both statistic models: i) the non-conditional variance is the same; ii) the non-conditional mean and the kurtosis are different; iii) the values of the asymmetry coefficient are negative; iv) the conditional mean *Type II* depends, for the MGARCH, on 1 period behind[14], while for the TAR, it depends on the last 6 periods; and v) the conditional variance *Type III* in the model MGARCH is in function to the variables $\sigma_{i,t-1}^2$ and $a_{i,t-1}$ ($i = 1,2$), while for the TAR, it depends only on $x_{t-j}$ ($j = 1,...,6$).

**Table 9 here**

The Figure 5 illustrates the functions of the conditional variance of both models, where it is observed the similarity of its dynamic behavior, thus, indicating that the TAR model manages to represent the conglomerates of external values[15].

**Figure 5 here**

### 4.5 Analysis of the leverage effect for the TAR and MGARCH models in the Bovespa case

Reminding that for the TAR model, the functional form of the NIC (equation(20)) is given by the volatility to the time $t$:

$$\sqrt[2]{Var(X_t|\tilde{x}_{t-1})_{TAR}} = [0.0002 + 0.5(0.0063 - 0.0478x^* - 0.0533x^* - 0.0779x^* - 0.0171x^* - 0.0508x^* - 0.0653x^*)^2 - (0.0002 - 0.0239x^* - \qquad (34)$$

---

[14] In this case, it is important to highlight that the model of the family ARCH depens both on the variabel $X_{t-1}$ and on $Z_{t-1}$, different to the threshold model that depends only on the past of the variable $\{X_t\}$.
[15] Similar fact to that found by Nieto and Moreno (2016) using Dow Jones' index as threshold variable.

$0.0267x^* - 0.0390x^* - 0.0086x^* - 0.0254x^* - 0.0327x^*)^2]^{\frac{1}{2}} = [0.0002 + 0.5(0.0063 - 0.3122x^*)^2 - (0.0002 - 0.1563x^*)^2]^{\frac{1}{2}}.$

Then, when calculating the equation (20) for determining the $x^*_{min}$, its value is equal to 0.0391, obtaining that the TAR model manages to capture the essence of the leverage effect, because of the new concept developed, $x^*_{min} > 0$. In the Figure 6, it is illustrated the graphic representation of the equation (34), where it is possible to observe that in values $x^*$ with the same magnitude but different sign, the volatility is higher when the sign is negative than when It is positive. For example, for an $x^* = -5\%$, the volatility is of 1.90%, while for an $x^* = 5\%$, it is of 1.31%.

**Figure 6 here**

For the MGARCH model, the NIS is used for representing graphically the leverage effect. In this case, assuming that $a_{2,t-1}$ takes the constant value equal to its historic average, the volatility to the time $t$ is

$$\sqrt[2]{Var(X_t|\tilde{x}_{t-1})_{MGARCH}} = [5.76 \times 10^{-6} + 0.0540 a_{1,t-1}^2 - 0.0812 a_{1,t-1} \overline{a_{2,t-1}} + 0.0305 \overline{a_{2,t-1}^2} + 0.8911 \overline{\sigma_{1,1,t-1}^2} + 0.0004 \overline{\sigma_{2,1,t-1}^2} + 4 \times 10^{-8} \overline{\sigma_{2,2,t-1}^2} + 0.0447 \zeta_{1,t-1}^2 + 0.0357 \zeta_{1,t-1} \overline{\zeta_{2,t-1}} + 0.0285 \overline{\zeta_{2,t-1}^2}]^{\frac{1}{2}}. \quad (35)$$

In the Figure 7, its graphic is shown, where like it occurs in the TAR model, it is obtained an asymmetric response in the volatility, clarifying that for this case the abscissa axis corresponds to the values of $a_{1,t-1}$, and not of $x^*$. Then, both graphics evidence the existence of the leverage effect in the Bovespa return series.

**Figure 7 here**

Following, the results of the estimations of the regressions proposed by French *et al.* (1987) are presented, corresponding to the case of the TAR and MGARCH models for the Bovespa logarithmic returns (*t* statistics in parentheses):

$$ln\left(\frac{\sigma_{t\,TAR}}{\sigma_{t-1\,TAR}}\right) = \underset{(0.2837)}{0.0001} - \underset{(-21.5693)}{0.7766}\, r_{t-1}. \tag{36}$$

$$ln\left(\frac{\sigma_{t\,MGARCH}}{\sigma_{t-1\,MGARCH}}\right) = \underset{(0.1540)}{0.0002} - \underset{(-27.4351)}{2.3865}\, r_{t-1}. \tag{37}$$

The coefficients that accompany the lagged return of a period ($\alpha_1$), of negative signs and statistically significant (to any level of common significance), confirm again the existence of a leverage effect, by finding a negative correlation between the rate of growth of the volatility and the past of the financial return.

## 5. Conclusions

As a first contribution of this research, it was managed to construct a theoretical approximation towards the representation of the news impact curves (NIC) from an open-loop threshold autoregressive (TAR) model, throughout the conditional marginal distribution to the past data of the interest variable. Using this, and under certain mathematic conditions, it is managed to characterize the leverage effect present in the financial series.

The second relevant contribution of this paper, comes from the empirical proof of a TAR model and its comparison against a model of the family of bivariate conditional heteroscedasticity (MGARCH), approximation never addressed before in the literature. Although the TAR model is not a system designed specifically for explaining the stylized facts of the financial series, it was found that it has an acceptable adjustment; in terms of asymmetry, dynamic of the functions of the conditional variance, and representation of the leverage effect.

As additional results, in this research, analytic expressions for the calculation of the asymmetry and the kurtosis of a TAR model stationary in a weak sense were found. In general terms, it was found that these two measures are in function of the expected value of $X_t$ conditioned to the regimen $j$, the non-conditional expected value of $X_t$ and the weights of each regimen $j$.

For future researches, it is proposed to study under which conditions over the parameters of a $TAR(l, k_1, k_2, \ldots, k_l)$ model, $x *$ is positive. Another aspect to delve, will be the possible extension of the news impact curve through Multivariate TAR (MTAR) stochastic processes.

32. Kawakatsu, H. (2006). Matrix exponential GARCH. *Journal of Econometrics* 134(1): 95–128.

33. Knapp, T. (2007). Bimodality revisited. *Journal of Modern Applied Statistical Methods* 6(1): 8–20.

34. Kodres, L., Narain, A. (2012). Fixing the system. *Finance & Development* 49(2): 14–16.

35. Koizumi, K., Okamoto, N., Seo, T. (2009). On Jarque-Bera tests for assessing multivariate normality. *Journal of Statistics: Advances in Theory and Applications* 1(2): 207–220.

36. Kroner, K., Ng, V. (1998). Modelling asymmetric comovements of asset returns. *The Review of Financial Studies* 11(4): 817–844.

37. Lütkepohl, H. (2005). *New introduction to multiple time series analysis.* Cambridge: Springer.

38. McAleer, M., Hoti, S., Chan, F. (2009). Structure and asymptotic theory for multivariate asymmetric conditional volatility. *Econometric Reviews* 28(5): 422–440.

39. Moreno, E. (2010). Una aplicación del modelo TAR en series de tiempo financieras (Tesis de Maestría). Universidad Nacional de Colombia, Bogotá, D.C.

40. Nieto, F. (2005). Modeling bivariate threshold autoregressive processes in the presence of missing data. *Communications in Statistics - Theory and Methods* 34(4): 905–930.

41. Nieto, F. (2008). Forecasting with univariate TAR models. *Statistical Methodology* 5(3): 263–276.

42. Nieto, F., Hoyos, M. (2011). Testing linearity against a univariate TAR specification in time series with missing data. *Revista Colombiana de Estadística* 34(1): 73–94.

43. Nieto, F., Moreno, E. (2014). Modelos TAR en series de tiempo financieras. *Comunicaciones en Estadística* 7(2): 223–243.

44. Nieto, F., Moreno, E. (2016). On the univariate conditional distributions of an open-loop TAR stochastic process. *Revista Colombiana de Estadística* 39(2): 149-165.

45. Nieto, F., Ruiz, F. (2002). About a prompt strategy for estimating missing data in long time series. *Revista de la Academia Colombiana de Ciencias Exactas, Físicas y Naturales* 26(100): 411–418.

**Address**

Corresponding author. E-mail: oaespinosaa@unal.edu.co

Carrera 30 # 45-03, Edificio 311, Deanery Office, Economic Sciences Faculty, Universidad Nacional de Colombia, Bogotá, Colombia.

**Appendix**

*Proof of the Proposition 1*

By assuming that the TAR stochastic process $\{X_t\}$ is stationary in a weak sense, initially, the asymmetry of $X_t$ can be defined as

$$\propto_3 = \frac{E(X_t - \mu)^3}{[E(X_t - \mu)^2]^{\frac{3}{2}}}, \qquad (38)$$

For all $t$. Or the model $TAR(l; k_1, \ldots, k_l)$ defined in (1), and knowing that the function of accumulated distribution of $X_t$ is given by

$$F_t(x) = \sum_{j=1}^{l} p_j \left[ F_{t,j}(x) \right],$$

where $F_{t,j}(x) = P(X_t \leq x | Z_t \in B_j)$, it is obtained

$$E(X_t - \mu)^3 = \int_{-\infty}^{\infty} (X - \mu)^3 dF_t(x) = \int_{-\infty}^{\infty} (X - \mu)^3 d \left( \sum_{j=1}^{l} p_j F_{t,j}(x) \right)$$

$$= \sum_{j=1}^{l} p_j \int_{-\infty}^{\infty} (X - \mu)^3 d(F_{t,j}(x)) = \sum_{j=1}^{l} p_j E\left[(X_t - \mu)^3 | Z_t \in B_j\right].$$

And by defining $\mu'_{j,3} = E\left[(X_t - \mu)^3 | Z_t \in B_j\right]$,

$$E(X_t - \mu)^3 = \sum_{j=1}^{l} p_j \mu'_{j,3}. \tag{39}$$

Besides, by defining $\mu'_{j,2} = E\left[(X_t - \mu)^2 | Z_t \in B_j\right]$, it is obtained the second central moment around the mean, necessary for computing the denominator of $\propto_3$. This is

$$E(X_t - \mu)^2 = \sum_{j=1}^{l} p_j \mu'_{j,2}. \tag{40}$$

So, replacing (39) and (40) in (38),

$$\propto_3 = \frac{\sum_{j=1}^{l} p_j \mu'_{j,3}}{\left( \sum_{j=1}^{l} p_j \mu'_{j,2} \right)^{3/2}}, \tag{41}$$

For each $t$.

Resuming the information presented in Table 1, where it is known that $X_t|Z_t \in B_j \sim N\left(\psi_j(1)\, a_0^{(j)}, (h^{(j)}\bar{\sigma}_j)^2\right)$, by means of the generating function of moments, it is possible to calculate the first three central moments of $X_t|Z_t \in B_j$ around zero, as follows[16]:

$$M_{X_t|Z_t \in B_j}(t) = e^{\psi_j(1)a_0^{(j)}t + \frac{1}{2}[h^{(j)}]^2 \bar{\sigma}_j^2 t^2},$$

$$M_{X_t|Z_t \in B_j}{}'(0) = \psi_j(1)a_0^{(j)},$$

$$M_{X_t|Z_t \in B_j}{}''(0) = [h^{(j)}]^2 \bar{\sigma}_j^2 + \left[\psi_j(1)a_0^{(j)}\right]^2,$$

then

$$M_{X_t|Z_t \in B_j}{}'''(t) = 3[h^{(j)}]^2 \bar{\sigma}_j^2 \left(\psi_j(1)a_0^{(j)} + [h^{(j)}]^2 \bar{\sigma}_j^2 t\right) e^{\psi_j(1)a_0^{(j)}t + \frac{1}{2}[h^{(j)}]^2 \bar{\sigma}_j^2 t^2}$$

$$+ \left(\psi_j(1)a_0^{(j)} + [h^{(j)}]^2 \bar{\sigma}_j^2 t\right)^3 e^{\psi_j(1)a_0^{(j)}t + \frac{1}{2}[h^{(j)}]^2 \bar{\sigma}_j^2 t^2},$$

and

$$M_{X_t|Z_t \in B_j}{}'''(0) = 3\psi_j(1)a_0^{(j)}[h^{(j)}]^2 \bar{\sigma}_j^2 + \left[\psi_j(1)a_0^{(j)}\right]^3.$$

Thus

$$\mu'_{t,j,3} = E(X_t^3|Z_t \in B_j) - 3\mu E(X_t^2|Z_t \in B_j) + 3\mu^2 E(X_t|Z_t \in B_j) - \mu^3$$

$$= 3\psi_j(1)a_0^{(j)}[h^{(j)}]^2 \bar{\sigma}_j^2 + \left[\psi_j(1)a_0^{(j)}\right]^3 - 3\mu[h^{(j)}]^2 \bar{\sigma}_j^2 - 3\mu\left[\psi_j(1)a_0^{(j)}\right]^2 + 3\mu^2 \psi_j(1)a_0^{(j)} - \mu^3.$$

By taking $\mu_{j,1} = \psi_j(1)a_0^{(j)}$ y $\sigma_j^2 = [h^{(j)}]^2 \bar{\sigma}_j^2$,

$$\mu'_{j,3} = 3\mu_{j,1}\sigma_j^2 + \mu_{j,1}^3 - 3\mu\sigma_j^2 - 3\mu\mu_{j,1}^2 + 3\mu^2\mu_{j,1} - \mu^3$$

---

[16] The apostrophe used in the following expressions makes reference to the order of the derivative.

$$= 3\mu_{j,1}\sigma_j^2 - 3\mu\sigma_j^2 + (\mu_{j,1} - \mu)^3$$

$$= 3\sigma_j^2(\mu_{j,1} - \mu) + (\mu_{j,1} - \mu)^3$$

$$= \underbrace{(\mu_{j,1} - \mu)}_{Ambiguous\ Sign}\underbrace{(3\sigma_j^2 + (\mu_{j,1} - \mu)^2)}_{Positive\ Sign(+)}.$$

And knowing that,

$$\mu'_{j,2} = E(X_t^2|Z_t \in B_j) - 2\mu E(X_t|Z_t \in B_j) + \mu^2$$

$$= [h^{(j)}]^2\bar{\sigma}_j^2 + [\psi_j(1)a_0^{(j)}]^2 - 2\mu\psi_j(1)a_0^{(j)} + \mu^2$$

$$= \sigma_j^2 + \mu_{j,1}^2 - 2\mu\mu_{j,1} + \mu^2,$$

the following analytic expression is obtained,

$$\alpha_3 = \frac{\sum_{j=1}^{l} p_j[(\mu_{j,1} - \mu)(3\sigma_j^2 + (\mu_{j,1} - \mu)^2)]}{[\sum_{j=1}^{l} p_j(\sigma_j^2 + (\mu_{j,1} - \mu)^2)]^{\frac{3}{2}}}, \qquad (42)$$

This completes the proof.

*Proof of the Proposition 2*

By satisfying the conditions given in (3) and assuming that the stochastic process $\{Z_t\}$ has identical univariate marginal distributions, the kurtosis of the interest variable $X_t$ can be defined as

$$\alpha_4 = \frac{E(X_t - \mu)^4}{[E(X_t - \mu)^2]^2}, \qquad (43)$$

for all $t$, where $\mu = E(X_t)$. For this reason, it is necessary to calculate the fourth derivative of the generating function of moments,

$$M_{X_t|Z_t\in B_j}''''(t) = 3[h^{(j)}]^4 \bar{\sigma}_j^4 e^{\psi_j(1)a_0^{(j)}t+\frac{1}{2}[h^{(j)}]^2\bar{\sigma}_j^2 t^2} + 6[h^{(j)}]^2 \bar{\sigma}_j^2 \left(\psi_j(1)a_0^{(j)} + [h^{(j)}]^2 \bar{\sigma}_j^2 t\right)^2 e^{\psi_j(1)a_0^{(j)}t+\frac{1}{2}[h^{(j)}]^2\bar{\sigma}_j^2 t^2} + \left(\psi_j(1)a_0^{(j)} + [h^{(j)}]^2 \bar{\sigma}_j^2 t\right)^4 e^{\psi_j(1)a_0^{(j)}t+\frac{1}{2}[h^{(j)}]^2\bar{\sigma}_j^2 t^2},$$

thus

$$M_{X_t|Z_t\in B_j}''''(0) = 3[h^{(j)}]^4 \bar{\sigma}_j^4 + 6[h^{(j)}]^2 \bar{\sigma}_j^2 \left[\psi_j(1)a_0^{(j)}\right]^2 + \left[\psi_j(1)a_0^{(j)}\right]^4.$$

Therefore

$$\mu'_{j,4} = E\big[(X_t - \mu)^4 | Z_t \in B_j\big]$$

$$= E(X_t^4 | Z_t \in B_j) - 4\mu E(X_t^3 | Z_t \in B_j) + 6\mu^2 E(X_t^2 | Z_t \in B_j) - 4\mu^3 E(X_t | Z_t \in B_j) + \mu^4$$

$$= 3\sigma_j^4 + 6\sigma_j^2 \mu_{j,1}^2 + \mu_{j,1}^4 - 12\mu_{j,1}\sigma_j^2 \mu - 4\mu_{j,1}^3 \mu + 6\sigma_j^2 \mu^2 + 6\mu_{j,1}^2 \mu^2 - 4\mu^3 \mu_{j,1} + \mu^4.$$

Reorganizing,

$$\mu'_{j,4} = \mu_{j,1}^4 - 4\mu_{j,1}^3 \mu + 6\mu_{j,1}^2 \mu^2 - 4\mu^3 \mu_{j,1} + \mu^4 + 6\sigma_j^2 \mu_{j,1}^2 - 12\sigma_j^2 \mu_{j,1}\mu + 6\sigma_j^2 \mu^2 + 3\sigma_j^4$$

$$= (\mu_{j,1} - \mu)^4 + 6\sigma_j^2 (\mu_{j,1} - \mu)^2 + 3\sigma_j^4,$$

and by factorizing and replacing in (43), besides using some calculations of the proof of the *proposition 1* it is obtained that,

$$\alpha_4 = \frac{\sum_{j=1}^{l} p_j \big[(\mu_{j,1} - \mu)^4 + 6\sigma_j^2 (\mu_{j,1} - \mu)^2 + 3\sigma_j^4\big]}{\big[\sum_{j=1}^{l} p_j \big(\sigma_j^2 + (\mu_{j,1} - \mu)^2\big)\big]^2}. \tag{44}$$

This completes the proof.

**Figure 1**. Bovespa index (a) and its returns (b) and S&P 500 index (c) and its returns (d).

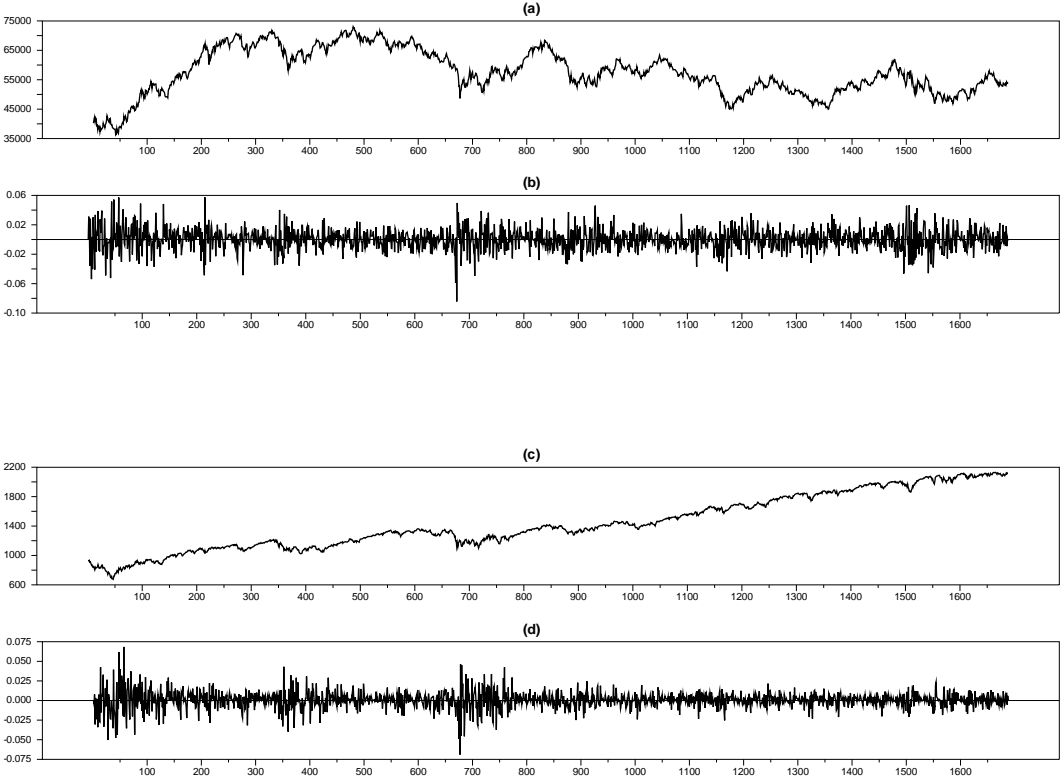

**Source**: Elaborated by the authors.

**Figure 2**. Empiric distribution of the Bovespa returns vs. Normal distribution simulation.

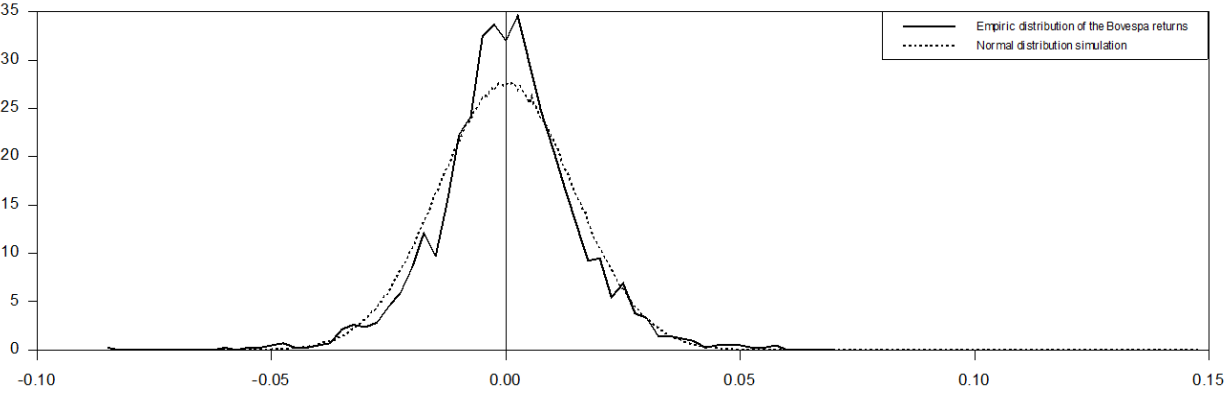

**Source**: Elaborated by the authors.

**Figure 3**: Autocorrelation function (a) and partial autocorrelation (b) of the standardized residuals of the TAR(2;0,6) model of Bovespa.

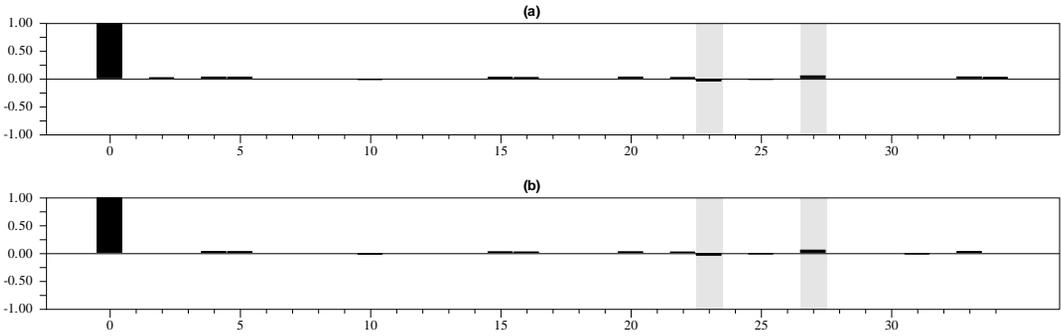

**Source**: Elaborated by the authors.

**Figure 4**: CUSUM (a) and CUSUMSQ (b) for the standardized residuals of the TAR(2;0,6) model of Bovespa.

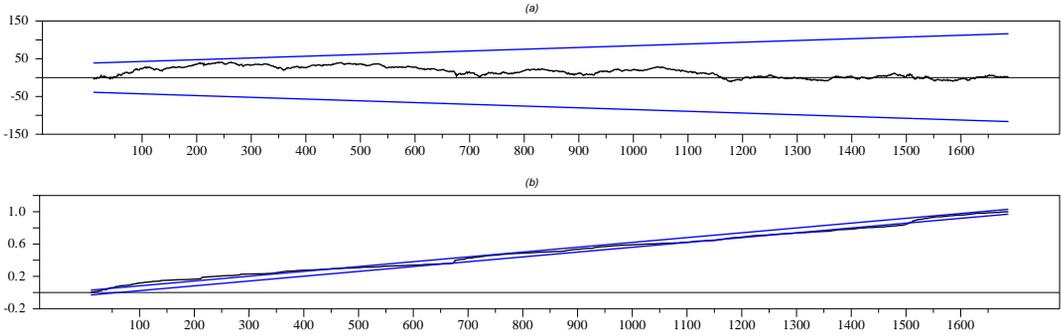

**Source**: Elaborated by the authors.

**Figure 5**: Functions of conditional variance estimated for the TAR(2;0,6) (a) and VAR(1)-A-BEKK(1,1) (b) models of Bovespa.

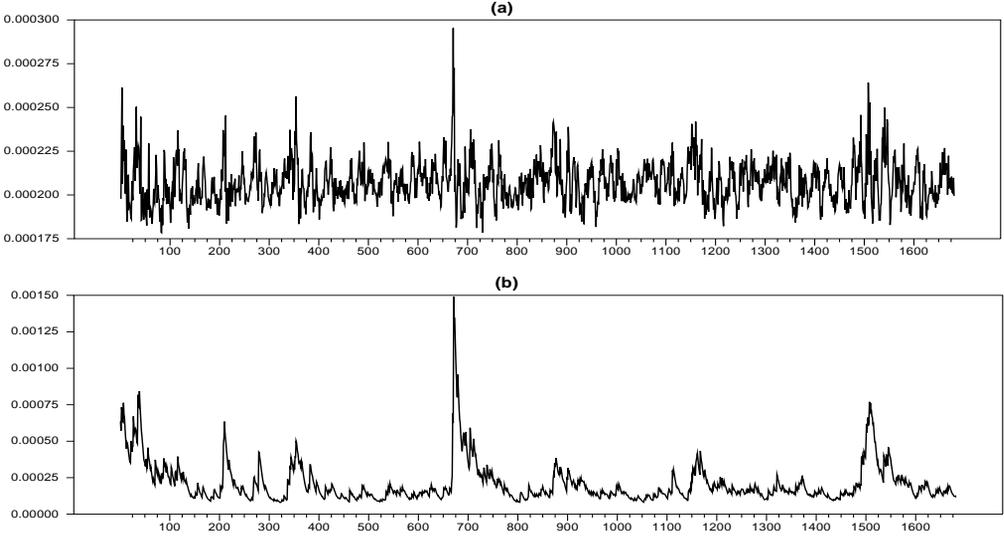

**Source**: Elaborated by the authors.

**Figure 6**: New information impact curve for the TAR(2;0,6) model of Bovespa.

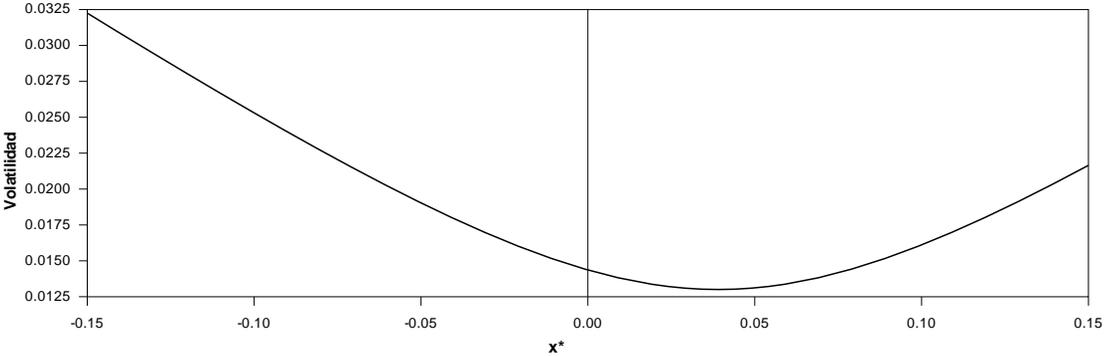

**Source**: Elaborated by the authors.

**Figure 7**: New information impact curve for the VAR(1)-A-BEKK(1,1) model of Bovespa.

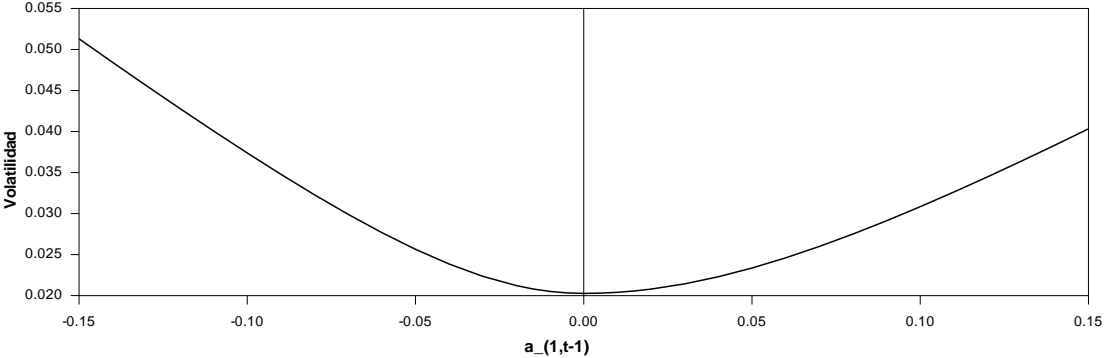

**Source**: Elaborated by the authors.

**Table 1.** Means and conditional variances of a TAR model ($l;\ k_1, k_2, \ldots, k_l$).

| Conditioned in | Mean | Variance |
|---|---|---|
| A regimen, $Z_t \in B_j$<br><br>**Type I** | $E(X_t \mid Z_t \in B_j) = \psi_j(1)\, a_0^{(j)}$,<br>where $\psi_j(1) = \dfrac{1}{\phi_j(1)} = \sum_{i=0}^{\infty} \psi_i^{(j)}$ | $Var(X_t \mid Z_t \in B_j) = \left(h^{(j)} \bar{\sigma}_j\right)^2$,<br>where $\bar{\sigma}_j = \sqrt[2]{\left(\sum_{i=0}^{\infty} \left(\psi_i^{(j)}\right)^2\right)}$ |
| A regimen, $Z_t \in B_j$,<br>and the past data,<br>$\tilde{x}_{t-1}$<br><br>**Type II** | $E(X_t \mid Z_t \in B_t, \tilde{x}_{t-1}) =$<br>$a_0^{(j)} + \sum_{i=1}^{k_j} a_i^{(j)} x_{t-i}$ | $Var(X_t \mid Z_t \in B_t, \tilde{x}_{t-1}) = \left(h^{(j)}\right)^2$ |
| The past data, $\tilde{x}_{t-1}$<br><br>*Type III* | $E(X_t \mid \tilde{x}_{t-1}) =$<br>$\sum_{j=1}^{l} p_j a_0^{(j)} + \sum_{j=1}^{l} p_j a_1^{(j)} x_{t-1} + \cdots$<br>$+ \sum_{j=1}^{l} p_j a_{k_j}^{(j)} x_{t-k_j}$ | $Var(X_t \mid \tilde{x}_{t-1}) =$<br>$\sum_{j=1}^{l} p_j \left(h^{(j)}\right)^2 + \sum_{j=1}^{l} p_j \mu_j^2$<br>$- \left(\sum_{j=1}^{l} p_j \mu_j\right)^2$,<br>where<br>$\mu_j = a_0^{(j)} + \sum_{i=1}^{k_j} a_i^{(j)} x_{t-i}$ |

**Source:** Elaborated by the authors.

**Table 2.** Results of the simulations corresponding to the asymmetry for the models M1, M2 and M3.

| Model | Coefficient of the asymmetry $\alpha_3$ | Mean of the sample asymmetry $\hat{\alpha}_3$ | Standard deviation of $\hat{\alpha}_3$ $ds_{\hat{\alpha}_3}$ | Interval $\hat{\alpha}_3 \pm 2 ds_{\hat{\alpha}_3}$ |
|---|---|---|---|---|
| **M1** | -0.1179 | -0.1626 | 0.2061 | (-0.5748, 0.2496) |
| **M2** | 0.2603 | 0.3446 | 0.1237 | (0.0972, 0.5920) |
| **M3** | -0.6558 | -0.7312 | 0.2232 | (-1.1776, -0.2849) |

**Source:** Elaborated by the authors.

**Table 3.** Results of the simulations corresponding to the kurtosis for the models M1, M2 and M3.

| Model | Coefficient of kurtosis $\alpha_4$ | Mean of the sample kurtosis $\hat{\alpha}_4$ | Standard deviation of $\hat{\alpha}_4$ $ds_{\hat{\alpha}_4}$ | Interval $\hat{\alpha}_4 \pm 2ds_{\hat{\alpha}_4}$ |
|---|---|---|---|---|
| **M1** | 3.2659 | 3.6046 | 0.4515 | (2.7016, 4.5076) |
| **M2** | 2.6847 | 2.7826 | 0.2669 | (2.2487, 3.3165) |
| **M3** | 3.6497 | 4.7848 | 0.7600 | (3.2648, 6.3047) |

**Source**: Elaborated by the authors.

**Table 4.** Estimated coefficients of the TAR(2;0,6) model of Bovespa.

| | **Regimen 1** | **Regimen 2** |
|---|---|---|
| $a_0^{(j)}$ | -0.0059 <br> (4.5500x10$^{-4}$) <br> [-0.0066, -0.0051] | 0.0063 <br> (4.4665x10$^{-4}$) <br> [0.0055, 0.0070] |
| $a_1^{(j)}$ | | -0.0478 <br> (0.0208) <br> [-0.08163, -0.01395] |
| $a_2^{(j)}$ | | -0.0533 <br> (0.0302) <br> [-0.1029, -0.0025] |
| $a_3^{(j)}$ | | -0.0779 <br> (0.0307) <br> [-0.1288, -0.0285] |
| $a_4^{(j)}$ | | -0.0171 <br> (0.0297) <br> [-0.0675, 0.0322] |
| $a_5^{(j)}$ | | -0.0508 <br> (0.0302) <br> [-0.1008, -0.0015] |
| $a_6^{(j)}$ | | -0.0653 <br> (0.0301) <br> [-0.1145, -0.0158] |
| $(h^{(j)})^2$ | 1.7418x10$^{-4}$ <br> (8.6104x10$^{-6}$) <br> [1.6060x10$^{-4}$, 1.8886x10$^{-4}$] | 1.6405x10$^{-4}$ <br> (8.0352x10$^{-6}$) <br> [1.5108x10$^{-4}$, 1.7764x10$^{-4}$] |

Typical deviations in parenthesis and credible intervals (90%) in square brackets.

**Source**: Elaborated by the authors.

**Table 5**: Results of the estimation of the VAR(1)-A-BEKK(1,1) model of Bovespa[17].

| | | | |
|---|---|---|---|
| **VAR(1) Model** | **μ, Γ** | $\mu_1$ | -2.4820x10$^{-5}$ (-0.0864) |
| | | $\Gamma_{1,1}$ | -0.0142 (-0.5069) |
| | | $\Gamma_{1,2}$ | 0.0425 (1.1631) |
| | | $\mu_2$ | 0.0004 (2.1686) |
| | | $\Gamma_{2,1}$ | 0.0449 (2.9989) |
| | | $\Gamma_{2,2}$ | -0.0578 (-2.1030) |
| **A-BEKK(1,1) Model** | ***C*** | $c_{1,1}$ | 0.0024 (8.0344) |
| | | $c_{1,2}$ | 0.0008 (3.9533) |
| | | $c_{2,2}$ | -0.0012 (-10.0756) |
| | **λ** | $\lambda_{1,1}$ | -0.2324 (-6.8033) |
| | | $\lambda_{1,2}$ | 0.0155 (1.9424) |
| | | $\lambda_{2,1}$ | 0.1746 (3.5402) |
| | | $\lambda_{2,2}$ | 0.0062 (0.1357) |
| | **ϑ** | $\vartheta_{1,1}$ | 0.9440 (87.9003) |
| | | $\vartheta_{1,2}$ | 0.0018 (0.2829) |
| | | $\vartheta_{2,1}$ | 0.0002 (0.0174) |

---

[17] For the estimation via maximum likelihood of the model MGARCH, the non-linear optimization method BFGS was used.

|  |  | $\vartheta_{2,2}$ | 0.9334 (98.7431) |
|---|---|---|---|
|  | **D** | $d_{1,1}$ | 0.2114 (3.9624) |
|  |  | $d_{1,2}$ | 0.0182 (1.9423) |
|  |  | $d_{2,1}$ | 0.1688 (2.4605) |
|  |  | $d_{2,2}$ | 0.4363 (11.6602) |
| *Joint contrasts for residuals* | | | |
|  | Lags | Statistic | P-value |
| **Q multivariate Test (Hosking, 1980; $H_0$: No Autocorrelation).** | 15 | 59.2439 | 0.5033 |
| **ARCH multivariate Test (Hacker and Hatemi-J, 2005; $H_0$: Homoscedasticity).** | 15 | 131.8612 | 0.5605 |
| **Jarque-Bera multivariate Test (Koizumi, Okamoto and Seo, 2009; $H_0$: Normality).** | - | 65.583 | 0.0000 |

*t* satistics in parentheses.

**Source**: Elaborated by the authors.

**Table 6**: Conditional moments on each regimen of the TAR(2;0,6) model of Bovespa.

|  | **Regimen 1** | **Regimen 2** |
|---|---|---|
| $\mu_{j,1}$ | -0.0059 | 0.0048 |
| $\mu_{j,2}$ | 0.0002 | 0.0002 |

**Source**: Elaborated by the authors.

**Table 7**: Conditional and non-conditional moments of the TAR(2;0,6) model of Bovespa.

| Moment | | Value | |
|---|---|---|---|
| Non-conditional mean | $E(X_t) = \sum_{j=1}^{l} p_j \mu_{j,1}$ | -0.0005 | |
| Non-conditional variance | $Var(X_t) = \sum_{j=1}^{l} p_j \mu_{j,2} - \left[\sum_{j=1}^{l} p_j \mu_{j,1}\right]^2$ | 0.0002 | |
| Non-conditional asymmetry | $\alpha_3$ | -0.0218 | |
| Non-conditional kurtosis | $\alpha_4$ | 2.9597 | |
| Conditional mean to the regimens | $E(X_t|Z_t \in B_j) = \psi_j(1) a_0^{(j)}$ | $j = 1$ | -0.0059 |
| | | $j = 2$ | 0.0048 |
| Conditional variance to the regimens | $Var(X_t|Z_t \in B_j) = \left(h^{(j)} \bar{\sigma}_j\right)^2$ | $j = 1$ | 0.0002 |
| | | $j = 2$ | 0.0002 |
| Conditional mean to the regimens and to the information until time $t-1$ | $E(X_t|Z_t \in B_j, \tilde{x}_{t-1})$ $= a_0^{(j)} + \sum_{i=1}^{k_j} a_i^{(j)} x_{t-i}$ | $j = 1$ | -0.0059 |
| | | $j = 2$ | 0.0063-0.0478$x_{t-1}$-0.0533$x_{t-2}$ -0.0779$x_{t-3}$-0.0171$x_{t-4}$ -0.0508$x_{t-5}$-0.0653$x_{t-6}$ |
| Conditional variance to the regimens and the information until time $t-1$ | $Var(X_t|Z_t \in B_j, \tilde{x}_{t-1}) = \left(h^{(j)}\right)^2$ | $j = 1$ | 0.0002 |
| | | $j = 2$ | 0.0002 |

| | | |
|---|---|---|
| **Conditional mean to the information until time $t-1$** | $E(X_t\|\tilde{x}_{t-1}) = \sum_{j=1}^{l} p_j \left( a_0^{(j)} + \sum_{i=1}^{k_j} a_i^{(j)} x_{t-i} \right)$ | $0.0002 - 0.0239 x_{t-1} - 0.0267 x_{t-2} - 0.0390 x_{t-3} - 0.0086 x_{t-4} - 0.0254 x_{t-5} - 0.0327 x_{t-6}$ |
| **Conditional variance to the information until time $t-1$** | $Var(X_t\|\tilde{x}_{t-1}) = \sum_{j=1}^{l} p_j (h^{(j)})^2 + \sum_{j=1}^{l} p_j \left( a_0^{(j)} + \sum_{i=1}^{k_j} a_i^{(j)} x_{t-i} \right)^2 - \left( \sum_{j=1}^{l} p_j \left( a_0^{(j)} + \sum_{i=1}^{k_j} a_i^{(j)} x_{t-i} \right) \right)^2$ | $0.0002 + 0.5(0.0063 - 0.0478 x_{t-1} - 0.0533 x_{t-2} - 0.0779 x_{t-3} - 0.0171 x_{t-4} - 0.0508 x_{t-5} - 0.0653 x_{t-6})^2 - (0.0002 - 0.0239 x_{t-1} - 0.0267 x_{t-2} - 0.0390 x_{t-3} - 0.0086 x_{t-4} - 0.0254 x_{t-5} - 0.0327 x_{t-6})^2$ |

**Source**: Elaborated by the authors.

**Table 8**: Conditional and non-conditional moments of the VAR(1)-A-BEKK(1,1)[18] model of Bovespa.

| Moment | Value |
|---|---|
| **Non-conditional mean** | $E(X_t)$=0.0002 |
| **Non-conditional variance** | $Var(X_t)$=0.0002 |
| **Non-conditional asymmetry** | $\alpha_3$=-0.0656 |
| **Non-conditional kurtosis** | $\alpha_4$=4.839 |
| **Conditional mean to the information until time $t-1$** | $E(X_t\|\tilde{x}_{t-1})$= -2.4820x10$^{-5}$ - 0.0142$X_{t-1}$+0.0425$Z_{t-1}$ |
| **Conditional variance to the information until time $t-1$** | $Var(X_t\|\tilde{x}_{t-1})$=5.76x10$^{-6}$ + 0.0540$a_{1,t-1}^2$ − 0.0812$a_{1,t-1}a_{2,t-1}$ + 0.0305$a_{2,t-1}^2$ + 0.8911$\sigma_{1,1,t-1}^2$ + 0.0004$\sigma_{2,1,t-1}^2$ + 4x10$^{-8}\sigma_{2,2,t-1}^2$ + 0.0447$\zeta_{1,t-1}^2$ + 0.0357$\zeta_{1,t-1}\zeta_{2,t-1}$ + 0.0285$\zeta_{2,t-1}^2$ |

**Source**: Elaborated by the authors.

---

[18] For the MGARCH model the value of the moments obtained by its residuals was taken as proxy (Caporin and McAleer, 2011).

**Table 9**: Comparison between the conditional and non-conditional moments of the TAR(2;0,6) and VAR(1)-A-BEKK(1,1) models of Bovespa.

| Moment | TAR | MGARCH |
|---|---|---|
| $E(X_t)$ | -0.0005 | 0.0002 |
| $Var(X_t)$ | 0.0002 | 0.0002 |
| $\alpha_3$ | -0.0218 | -0.0656 |
| $\alpha_4$ | 2.9597 | 4.839 |
| $E(X_t\|\tilde{x}_{t-1})$ | 0.0002-0.0239$x_{t-1}$-0.0267$x_{t-2}$ -0.0390$x_{t-3}$-0.0086$x_{t-4}$-0.0254$x_{t-5}$ -0.0327$x_{t-6}$ | -2.4820x10$^{-5}$ - 0.0142$x_{t-1}$+0.0425$z_{t-1}$ |
| $Var(X_t\|\tilde{x}_{t-1})$ | 0.0002+0.5(0.0063-0.0478$x_{t-1}$-0.0533$x_{t-2}$-0.0779$x_{t-3}$-0.0171$x_{t-4}$-0.0508$x_{t-5}$-0.0653$x_{t-6}$)$^2$-(0.0002-0.0239$x_{t-1}$-0.0267$x_{t-2}$-0.0390$x_{t-3}$-0.0086$x_{t-4}$-0.0254$x_{t-5}$-0.0327$x_{t-6}$)$^2$ | $=5.76 \times 10^{-6} + 0.0540 a_{1,t-1}^2 - 0.0812 a_{1,t-1} a_{2,t-1} + 0.0305 a_{2,t-1}^2 + 0.8911 \sigma_{1,1,t-1}^2 + 0.0004 \sigma_{2,1,t-1}^2 + 4 \times 10^{-8} \sigma_{2,2,t-1}^2 + 0.0447 \zeta_{1,t-1}^2 + 0.0357 \zeta_{1,t-1} \zeta_{2,t-1} + 0.0285 \zeta_{2,t-1}^2$ |

**Source**: Elaborated by the authors.